\documentclass[final,5p,times]{elsarticle}

\usepackage{amssymb}
\usepackage{lineno}
\usepackage{url}

\newcommand\E{\mathop{\mathit{E}}\nolimits}
\newcommand\var{\mathop{\mathit{Var}}\nolimits}
\newcommand\cov{\mathop{\mathit{Cov}}\nolimits}
\newcommand\mse{\mathop{\mathit{MSE}}\nolimits}
\newcommand\diag{\mathop{\mathrm{diag}}\nolimits}

\newcommand\blue{\mathop{\mathit{BLUE}}\nolimits}
\newcommand\blup{\mathop{\mathit{BLUP}}\nolimits}

\newcommand\ml{\mathop{\mathit{ML}}\nolimits}
\newcommand\reml{\mathop{\mathit{REML}}\nolimits}
\newcommand\tr{\mathop{\mathrm{tr}}\nolimits}
\newcommand\rank{\mathop{\mathrm{rank}}\nolimits}
\newcommand\loglik{\mathop{\mathrm{loglik}}\nolimits}

\journal{Measurement Science Review, Vol.~12, No.~6, 2012, 234--248.}

\begin{document}
\begin{frontmatter}

\title{Estimation, Testing, and Prediction Regions of the Fixed and Random Effects \\ by Solving the Henderson's Mixed Model Equations}
\author[um]{Viktor Witkovsk\'y\corref{cor1}}
\ead{witkovsky@savba.sk}
\cortext[cor1]{Corresponding author. Tel.: +421 2 59104530; Fax: +421 2 54775943.}
\address[um]{Institute of Measurement Science, Slovak Academy of Sciences, Bratislava, Slovakia}

\begin{abstract}
We present a brief overview of the methods for making statistical inference (testing statistical hypotheses, construction of confidence and/or prediction intervals and regions) about linear functions of the fixed effects and/or about the fixed and random effects simultaneously, in conventional simple linear mixed model. The presented approach is based on solutions from the Henderson's mixed model equations.
\end{abstract}


\end{frontmatter}


\section{Introduction}
The applications of data analysis based on the statistical linear mixed model, as a natural generalization of the analysis of variance methods and the ANOVA models, (see e.g.~\cite{Robinson1991}, \cite{Gelman2005}, \cite{McCulloch2001}),  are widespread. Such applications with analytical methods based on linear mixed models include different fields of the biomedical and technical research, (see \cite{Volaufova2005} and/or \cite{Domotor2012}). For illustration, here we shall mention just few of them: e.g.~genetics with its microarray experiments, \cite{Cui2003}, \cite{Cui2005}, \cite{Cui2006}, \cite{Wu2012}, the plant and animal breeding in agricultural, \cite{Calinski2008}, statistical meta-analysis in medical research, \cite{Hartung2008}, neurophysiology, \cite{Schulz2012},  as well as different technical applications, like e.g.~calibration of devices,  derivation of the tolerance intervals for industrial applications, interlaboratory comparisons in metrology, and methods for expression the uncertainties in measurements, see e.g.~\cite{Chvostekova2009}, \cite{Fonseca2007}, \cite{Herdahl2008}, \cite{Krishnamoorthy2009}, \cite{Savin2003}, \cite{Tucek2012}, \cite{Wimmer2007a}, \cite{Wimmer2007b}, \cite{Wimmer2011}, \cite{Witkovsky2005}, \cite{Witkovsky2003b}, \cite{Witkovsky2001b}, \cite{Witkovsky2003}, and \cite{Witkovsky2007}. 

Although the linear mixed models and the methods for statistical inference based on such models have been recognized and used for long time by the researchers in different fields, it seems that some sort of misunderstanding of the principles and/or the technical details (of the used methods for statistical inference based on such linear mixed models) may lead to improper usage of the implemented methods and algorithms. Moreover, there are still some further open theoretical problems (like e.g.~methods for testing and constructing confidence intervals/regions about the variance components, see e.g.~\cite{Arendacka2007}, \cite{Arendacka2012a}, \cite{Arendacka2012b}, \cite{Sirkova2001}, \cite{Volaufova1992a}, \cite{Volaufova1992b}, \cite{Volaufova2012}, \cite{Wimmer2003}, \cite{Witkovsky1996}, \cite{Witkovsky1998a}, \cite{Witkovsky1998b}).

So, the main goal of the paper is to present a brief overview of the standard (conventionally used) methods for making statistical inference (in particular the methods for testing statistical hypotheses and the methods for construction of the confidence and/or prediction intervals/regions) about linear functions of the fixed effects and/or about the fixed and random effects simultaneously, in conventional simple linear mixed model, (with pointing to potential problems which may appear based on usage of these methods),  and to present some of the  recently developed improvements, as well as some generalizations, together with relatively detailed technical description of the model and the methods. The presented approach is based on the elements of the solution of the Henderson's mixed model equations.

{
\section{Henderson's mixed model equations}\label{MMEsection}
We consider the  linear mixed model (LMM) in the following form
\begin{equation}\label{model1}
y=Xb + Zu + e,
\end{equation}
with $y$ being an $n$-dimensional vector of observations, $b$ being the $p$-vector of fixed effects, $u$ being the $r$-vector of random effects with $\E(u) = 0$ and $\var(u) = G$, and $e$ being the  $n$-vector of random (measurement) errors with $\E(e) = 0$ and $\var(e) = R$, where $R$ is assumed to be strictly positive definite variance-covariance matrix of $e$. The $(n\times p)$-matrix $X$ and the $(n\times r)$-matrix $Z$ are the known design matrices.
Typically, we can write $Zu=\sum_{i=1}^s Z_i u_i$, where the $(n\times r_i)$ matrices $Z_i$ and the $r_i$-dimensional random effects $u_i$, $i=1,\dots,s$, could be specified from the structure of the model.

The main goal of this paper is to present an overview of the methods for making statistical inference about linear functions of the fixed effects $b$ and the random effects $u$, i.e.~about $K'b$ and/or about $w = \Lambda'(b',u')' = K'b+L'u$ for given (suitable) coefficient matrices $\Lambda$, resp.~$K$ and $L$.

Henderson in \cite{Henderson1953} developed a set of equations, termed as the  mixed model equations (MMEs), that simultaneously yield the best linear unbiased estimator (BLUE)
of $Xb$ (or any vector of estimable linear functions $K'b$) and the best linear unbiased predictor (BLUP) of $u$ (or any
vector $w = K'b+L'u$, provided $K'b$ is estimable), under the assumption that the covariance structure is known.

The MMEs were derived based on the normality assumptions, i.e.~$u\sim N(0,G)$, $e \sim N(0,R)$, with $\cov(u,e)=0$, for known 
variance-covariance matrices $G$ and $R$. Thus, the joint probability density function (pdf) of the random vector
$(y',u')'$ is given as
{\small \begin{eqnarray}
\lefteqn{f(y,u)=f(y|u)f(u)}\cr
&&\frac{1}{(2\pi)^{n/2}|R|^{1/2}}\exp\left\{
-\frac 12 (y-Xb-Zu)'R^{-1}(y-Xb-Zu)\right\}\cr
&&\times
\frac{1}{(2\pi)^{r/2}|G|^{1/2}}\exp\left\{ - \frac 12
u'G^{-1}u\right\}.
\end{eqnarray}}
By solving the ML equations for $b$ and $u$, i.e.
\begin{equation}
\frac{\partial f(y,u)}{\partial b}=0, \qquad
\frac{\partial f(y,u)}{\partial u}=0
\end{equation}
we get  the MMEs in the following form
\begin{equation}\label{MME1}
\left(\begin{array}{cc}
X'R^{-1}X & X'R^{-1}Z\cr
Z'R^{-1}X & Z'R^{-1}Z+G^{-1}\end{array}\right)
\left(\begin{array}{c}\tilde{b} \cr\tilde{u} \end{array} \right)=
\left(\begin{array}{c}X'R^{-1}y\cr Z'R^{-1}y \end{array} \right).
\end{equation}
The left-hand side matrix of (\ref{MME1}) will be termed as the Henderson's MME matrix, here denoted by $H$, i.e.
\begin{equation}\label{H}
	H = (X,Z)'R^{-1}(X,Z) + (0,I_r)'G^{-1}(0,I_r),
\end{equation}
where by $0$ we denote a zero matrix with suitable dimensions, here $(r\times p)$.
Alternatively,
\begin{equation}\label{MME2}
\left(\begin{array}{cc}
X'R^{-1}X & X'R^{-1}ZG\cr
Z'R^{-1}X & W^{-1}\end{array}\right)
\left(\begin{array}{c}\tilde{b} \cr\tilde{v} \end{array} \right)=
\left(\begin{array}{c}X'R^{-1}y\cr Z'R^{-1}y \end{array} \right).
\end{equation}
where $W=(I+Z'R^{-1}ZG)^{-1} $.
Notice, that based on (\ref{MME2}), there is no need to restrict the variance-covariance matrix $G$ to be strictly positive definite. This version of MMEs is preferred for numerical evaluations, if $G$ can be a bad conditioned matrix.

Given the variance-covariance matrices $G$ and $R$, let us denote as $C$ the following matrix of coefficients
\begin{eqnarray}\label{C}
C &=&\left(\begin{array}{cc}
C_{11} & C_{12}\cr C_{21} & C_{22}
\end{array}\right)\cr
&=&\left(\begin{array}{cc}
X'R^{-1}X & X'R^{-1}Z\cr
Z'R^{-1}X & Z'R^{-1}Z+G^{-1}\end{array}\right)^{-}\cr
&=&\left(\begin{array}{cc} I_{p} & 0\cr
0&G\end{array}\right)
\left(\begin{array}{cc}
X'R^{-1}X & X'R^{-1}ZG\cr
Z'R^{-1}X & W^{-1}\end{array}\right)^{-},
\end{eqnarray}
where by $A^{-}$ we denote any $g$-inverse of the matrix $A$.

Let $\tilde{b}$ and  $\tilde{u}$  be any solution to the MMEs (\ref{MME1}). Notice that based on $\tilde{b}$ and  $\tilde{v}$, the solutions from (\ref{MME2}), we can reconstruct $\tilde{u}$ by $\tilde{u}=G\tilde{v}$.
Then the BLUE of the vector of linear estimable functions of the fixed effects $K'b$, see e.g.~\cite{Searle1992}, is
\begin{equation}
\blue(K'b)=K'\left(X'V^{-1}X\right)^-X'V^{-1}y=K'\tilde{b},
\end{equation}
where $K'$ is a $(q \times p)$-matrix of coefficients of the estimable linear function $K'b$, i.e.~$K=X'A$ for some matrix $A$, and
$V = Z'GZ + R$. 
The BLUP of the vector of linear functions of the fixed and random effects, say $K'b + L'u$,  is
\begin{eqnarray}
\blup(K'b+L'u)&=& \blue(K'b)\cr
&& +L'GZ'V^{-1}(y-\blue(Xb)),\cr
&=& K'\tilde{b}+L'\tilde{u},
\end{eqnarray}
where $L'$ is an arbitrary $(q \times r)$-matrix of coefficients, and $\blue(Xb)=X\tilde{b}$. 

Important properties of the solutions of the MMEs  are  summarized bellow, for more details see e.g.~\cite{McLean1991}:

\begin{enumerate}
\renewcommand{\itemsep}{-1pt}
\item
In the class of linear unbiased predictors, BLUP maximizes the correlation between $u$ and $\tilde{u}$.
\item
$K'\tilde{b}$ is BLUE of the set of estimable linear functions $K'b$.
\item
$\E\left(u \,|\, \tilde{u}\right) = \tilde{u}$.
\item
$\tilde{u}$ is unique.
\item
$K'\tilde{b} + L'\tilde{u}$ is BLUP of $K'b + L'u$ provided that $K'b$ is estimable.
\item
$\var\left(K'\tilde{b}\right) = K'C_{11}K$.
\item
$\var\left(K'\tilde{b} + L'\tilde{u}\right) = K'C_{11}K+ L'(G-C_{22})L$.
\item\label{prop}
$ \var\left(\left(K'\tilde{b}+L'\tilde{u}\right)- \left(K'b+L'u\right)\right) = (K',L')C (K',L')'$.
\item
$\cov\left(K'\tilde{b},\tilde{u}'\right)=0$.
\item
$\cov\left(K'\tilde{b}, u'\right)=-K'C_{12}$.
\item
$\cov\left(K'\tilde{b}, u'-\tilde{u}'\right)=-K'C_{12}$.
\item
$\var\left(\tilde{u}\right)=\cov\left(\tilde{u},u'\right)=G-C_{22}$.
\item
$\var\left(\tilde{u}-u\right)=C_{22}$.
\end{enumerate}

In this paper  we shall consider only a special form of the model (\ref{model1}) --- a conventional simple LMM with normally distributed errors and random effects. That is, we shall assume mutually uncorrelated (independent) normally distributed random effects $u_1,\dots,u_s$ and $e$ with
$\E(u_i)=0$ for $i=1,\dots,r$, $\E(e)=0$, $\cov(u_i,u_j)=0$ for $i\neq j$, and $\cov(u_i,e)=0$ for all $i=1,\dots,s$. 
Further, we shall assume $\var(u_i)=\sigma^2_i I_{r_i}$, $i=1,\dots,s$, with
$r=\sum_{i=1}^s r_i$, and $\var(e)=\sigma^2_{s+1} I_n$.
Hence,
\begin{equation}\label{model2}
\E(y) = Xb,\ \mathrm{and}\  \var(y) = \sum_{i=1}^s\sigma^2_i Z_iZ_i'  + \sigma^2_{s+1} I_n,
\end{equation}
with $\sigma^2=\left(\sigma^2_1,\dots,\sigma^2_s,\sigma^2_{s+1}\right)'$ being the vector of variance components with the parameter space specified by
$\sigma^2_i\geq0$ for  $i=1,\dots,s$, and $\sigma^2_{s+1}>0$. 
However, in order to avoid possible technical and numerical problems, it is reasonable to assume that the true parameter $\sigma^2=\left(\sigma^2_1,\dots,\sigma^2_s,\sigma^2_{s+1}\right)'$ is in the interior of this parameter space. So, here we shall assume that $\sigma^2_i>0$ for  $i=1,\dots,s+1$, 

In other words, we shall assume $y\sim N(Xb,V)$, with $V=\var(y)=ZGZ'+R$, where
$G$ is $(r\times r)$ diagonal matrix, $G=\var(u)=\diag(\sigma^2_i I_{r_i})$, and $R$ is $(n\times n)$ diagonal matrix, $R=\var(e)=\sigma^2_{s+1} I_n$, with $\sigma^2_i>0$ for  $i=1,\dots,s+1$.

If the variance components $\sigma^2=\left(\sigma^2_1,\dots,\sigma^2_s,\sigma^2_{s+1}\right)'$ are unknown, they can be (and in general must be) estimated from the observed data by any reasonably effective and computationally efficient method, like e.g.~by the methods based on moments (the minimum variance (norm) quadratic estimation) or the methods based on likelihood function (ML or REML).

There are several efficient implementations for estimation of the variance components in general LMMs.
One method used to fit such LMMs is the expectation-maximization (EM) algorithm, see~\cite{Lindstrom1988}, where the variance components are treated as unobserved nuisance parameters in the joint likelihood. 
Currently, such methods are implemented in the major statistical software packages SAS (Proc MIXED) and \texttt{R} (\texttt{lme} in the \texttt{nlme} library). 
In particular, Proc MIXED uses a ridge-stabilized Newton-Raphson algorithm to optimize either a full (ML) or residual (REML) likelihood function, see also~\cite{SAS}, \cite{Littel2006}, \cite{Wolfinger1994}, and \cite{Pinheiro2000}. 

However, here we present a relatively simple method, based on repeated iterative solving of the MMEs, suggested by Searle, Casella and McCulloch in \cite{Searle1992}.  The elements of MMEs are used for setting up iterative procedures for simultaneous estimation of the variance components $\sigma^2_1,\dots,\sigma^2_s,\sigma^2_{s+1}$ and the empirical versions of the BLUE of $b$ and the BLUP of $u$, in the simple LMM (\ref{model2}). 

The  algorithm provides  solution to the maximum likelihood (ML) or the restricted maximum likelihood (REML) equations for estimating variance components, see e.g.~\cite{Hartley1967},  \cite{Patterson1971}, \cite{Harville1977}, \cite{Laird1987}, and \cite{Searle1992}. The algorithm can be also  used for estimation of the related Fisher information matrices for ML and/or REML estimators of the variance components (i.e.~the inverse of the asymptotic variance-covariance matrix of the ML/REML estimators). Moreover, it can be also used  for computing the minimum norm quadratic estimates MINQE(I) (realizations of the invariant minimum norm quadratic estimators) or the  MINQE(U,I)  (invariant and unbiased minimum norm quadratic estimators) of the variance components, for more details see e.g.~\cite{LaMotte1973}, \cite{Rao1972}, and \cite{Kleffe1988}.

The final solutions of such iterative procedure will be denoted by $\hat{b}$, $\hat{u}=(\hat{u}_1',\dots,\hat{u}_{s}')'$, and $\hat{\sigma}^2= (\hat{\sigma}^2_1,\dots,\hat{\sigma}^2_{s+1})'$. 
Similarly, we shall use the adequate  notation $\hat{G}$, $\hat{R}$,  and $\hat{C}$ for the estimated versions of matrices $G$, $R$,  and $C$.
The solutions $\hat{b}$ and $\hat{u}$ satisfy the MMEs (\ref{MME1}) if the unknown matrices $G$ and $R$ are replaced by the estimated versions $\hat{G}$ and $\hat{R}$.
Finally, based on $\hat{\sigma}^2$, the important output of the algorithm is the estimated Fisher information matrix, say $I_{\ml}(\hat{\sigma}^2)$ or $I_{\reml}(\hat{\sigma}^2)$, respectively. Consequently, it provides the estimated asymptotic variance-covariance matrix of the estimated variance components $\hat{\sigma}^2$, say $\hat{\Sigma} =\left( I_{\ml}(\hat{\sigma}^2)\right)^{-1}$ or $\hat{\Sigma} =\left( I_{\reml}(\hat{\sigma}^2)\right)^{-1}$, provided that the inverses do exist. For detailed description of the algorithm see Section~\ref{Estimation}.
}

{
\section{Standard methods for statistical inference on fixed and random effects}
Here we consider the problem of making statistical inference about $q$ linear functions of the fixed effects $b$ and the random effects $u$, i.e.~about $\Lambda'\left(b',u'\right)'=K'b + L'u$
where $\Lambda$ is $((p+r)\times q)$-dimensional full-ranked matrix with estimable $K'b$ (i.e.~$K = X'A$ for some matrix $A$). 

Let $\tilde{b}$ and $\tilde{u}$ are the solutions of the MMEs (\ref{MME1}), so $\tilde{w} = \Lambda'\left(\tilde{b}',\tilde{u}'\right)' = K'\tilde{b} + L'\tilde{u}$ is the best linear unbiased predictor (BLUP) of $w = K'b + L'u$. Then, according to the properties 6 and 8 of Section~\ref{MMEsection}, the variance of $K'\tilde{b}$ and the  mean squared error (MSE) of $\tilde{w}$  are given by 
\begin{equation}\label{varKb}
\var(K'\tilde{b})	=  K'C_{11}K ,
\end{equation}
and
\begin{eqnarray}\label{MSE}
\mse\left(\tilde{w}\right) &=& E\left(\left(\tilde{w} -w \right)\left(\tilde{w} -w \right)'\right)\cr
&=& \var\left(\tilde{w} -w \right) = \Lambda'C\Lambda = M_{\tilde{w}}.
\end{eqnarray}
Notice that the MSE matrix of $\tilde{w}$, $M_{\tilde{w}}$, functionally depends on the variance components $\sigma^2=\left(\sigma^2_1,\dots,\sigma^2_s,\sigma^2_{s+1}\right)'$.

If the variance components $\sigma^2=\left(\sigma^2_1,\dots,\sigma^2_s,\sigma^2_{s+1}\right)'$ are known,  based on the model assumptions and from (\ref{varKb}) and (\ref{MSE}), we trivially get the pivot, Wald-type statistic, useful for making statistical inference about $K'b$ (e.g.~testing a null hypothesis $H_0: K'b = K'b_0$ for some $b_0$) and/or about the variable $w = K'b +L'u$ with their exact (null) distribution:
\begin{equation}
	Q = \left(K'\tilde{b}-K'b_0\right)'\left(K'C_{11} K \right)^{-1}\left(K'\tilde{b}-K'b_0\right) \sim \chi^2_q,
\end{equation}
and
\begin{equation}
	Q = \left(\tilde{w} -w \right)'\left(\Lambda'C\Lambda\right)^{-1}\left(\tilde{w} -w \right) \sim \chi^2_{q},
\end{equation}
where $\chi^2_{q}$ denotes the chi-squared distribution with $q = \rank(K') = \rank(\Lambda')$ degrees of freedom.

If the variance components are unknown and the estimated values $\hat{\sigma}^2= \left(\hat{\sigma}^2_1,\dots,\hat{\sigma}^2_{s+1}\right)'$ are available together with $\hat{C}$, a commonly used test statistic for fixed effects hypothesis $H_0: K'b = K'b_0$,  is based on $K'\hat{b}$ and $\hat{C}_{11}$:
\begin{equation}\label{F1}
	F = \frac{1}{q} \left(K'\hat{b}-K'b_0\right)'\left(K'\hat{C}_{11} K \right)^{-1}\left(K'\hat{b}-K'b_0\right),
\end{equation}
where $K'\hat{b}$ denotes the empirical version of the best linear unbiased estimator $K'\tilde{b}$  of $K'b$ (i.e.~version with the estimated variance-covariance components). Notice that $C_{11} = \left(X'V^{-1}X\right)^{-}$, see e.g.~\cite{Searle1992} (Eqn.~(55) p.~276),  and consequently $\hat{C}_{11} = \left(X'\hat{V}^{-1}X\right)^{-}$, where $\hat{V} = Z\hat{G}Z'+ \hat{R}$.

As a generalization, for making simultaneous statistical inference on the fixed as well as the random effects, i.e.~on $w =\Lambda'\left(b',u'\right)'$ (e.g.~construction of the prediction region) based on the empirical BLUP (EBLUP), i.e.~the predictor $\hat{w} = \Lambda'\left(\hat{b}',\hat{u}'\right)'$ (where $\hat{b}$ and $\hat{u}$ are solutions of the MMEs with estimated $\hat{R}$ and $\hat{G}$),  it is natural to consider the following statistic
\begin{equation}\label{F3}
F = \frac{1}{q} \left(\hat{w} -w \right)'\left(\Lambda'\hat{C}\Lambda\right)^{-1}\left(\hat{w} -w \right),
\end{equation}
where $q$ is rank of the matrix $\Lambda'$. 

As a special case, if $w$ is a one-dimensional function given by  $w = \lambda'\left(b',u'\right)' =  k'b +l'u$, in analogy with (\ref{F1}) and (\ref{F3}), it is natural to consider the pivot statistic
\begin{equation}\label{t1}
	t = \frac{k'\hat{b}-k'b_0}{\sqrt{k'\hat{C}_{11} k} } ,
\end{equation}
and/or its generalization
\begin{equation}\label{t3}
	t = \frac{\hat{w}-w}{\sqrt{\lambda' \hat{C}\lambda}} ,
\end{equation}
where $\hat{w} = \lambda'\left(\hat{b}',\hat{u}'\right)'$ is the EBLUP of $w$.

The (null) distribution of the statistics (\ref{t1}) and (\ref{t3}) is commonly approximated by the Student's $t$-distribution with $\nu$ degrees of freedom (DF), estimated by applying the Satterthwaite's approximation. The (null) distribution of the statistics (\ref{F1}) and (\ref{F3}) is commonly approximated by the Fisher-Snedecor's $F$-distribution with $\nu_1$ and $\nu_2$ degrees of freedom, where $\nu_1 = q$ and $\nu_2$, the denominator degrees of freedom (DDF), where $\nu_2$ is typically estimated by a generalization of the Satterthwaite's method, as suggested e.g.~by Fai and Cornelius in \cite{Fai1996}, or alternatively, by applying moment based approximation for the $F$-distribution. The explicit expressions for DF and DDF estimators of (\ref{t1}), (\ref{t3}), (\ref{F1}) and (\ref{F3}) are given in Sections~\ref{SectionSatt} and \ref{SectionFai}.

\subsection{DF estimated by the Satterthwaite's method}\label{SectionSatt}
Giesbrecht and Burns in \cite{Giesbrecht1985}, (see also \cite{McLean1988}, \cite{Elston1988}, and \cite{Schaalje2001}), suggested to approximate the null distribution of the
pivotal quantity (\ref{t1}) by the Student's $t$-distribution with $\hat{\nu}$ degrees of freedom (DF), where $\hat{\nu}$ is the Satterthwaite's approximation\footnote{The Satterthwaite's approximation of the distribution of $k'\hat{C}_{11}k$ is based on assumption that $\nu\left(k'\hat{C}_{11}k\right)/\sigma^2\sim\chi^2_\nu$ for some parameters $\sigma^2$ and $\nu$. By comparing the first and the second moments of both random variables we get $\E\left( \nu\left(k'\hat{C}_{11}k\right)/\sigma^2\right) = \nu$ and $\var\left( \nu\left(k'\hat{C}_{11}k\right)/\sigma^2\right) =  2\nu$. From that we directly get   $\sigma^2 = \E\left(k'\hat{C}_{11}k\right)$ and $\nu = 2\left(\E\left(k'\hat{C}_{11}k\right)\right)^2/\var\left(k'\hat{C}_{11}k\right)$.  As $\E\left(k'\hat{C}_{11}k\right)$ and $\var\left(k'\hat{C}_{11}k\right)$ depend on unknown parameters they should be estimated. So, we get the natural estimator as $\hat{\nu} = 2 \left(k'\hat{C}_{11}k\right)/\widehat{\var}\left(k'\hat{C}_{11}k\right)$.} of the (unknown) $\nu$, see \cite{Satterthwaite1941}, \cite{Satterthwaite1946}, i.e.
\begin{equation}\label{distt1}
	t = \frac{k'\hat{b}-k'b_0}{\sqrt{k'\hat{C}_{11} k} } \sim t_{\hat{\nu}_{k}},
\end{equation}
with
\begin{equation}\label{nu1}
	\hat{\nu}_{k} = \frac{2 \left(k'\hat{C}_{11} k\right)^2}{\widehat{\var}\left(k'\hat{C}_{11} k\right)} \equiv  \frac{2 \left(k'\hat{C}_{11} k\right)^2}{\hat{g}_{k}'\hat{\Sigma} \hat{g}_{k}},
\end{equation}
where $\widehat{\var}\left(k'\hat{C}_{11} k\right)$ denotes the estimated value of $\var\left(k'\hat{C}_{11} k\right)$.

The suggested estimator of $\widehat{\var}\left(k'\hat{C}_{11} k\right)\equiv \hat{g}_{k}'\hat{\Sigma} \hat{g}_{k}$ is based on the estimated version of the Taylor series expansion of the variance of the estimator $k'\tilde{b}$ (BLUE), i.e.~$\var\left(k'\tilde{b}\right) = k'C_{11}k$, with respect to the variance components $\sigma^2 = (\sigma^2_1,\dots,\sigma^2_s,\sigma^2_{s+1})$. Here, $\hat{\Sigma}$ is the estimated (asymptotic) variance-covariance matrix of the estimators (e.g.~REML estimators) of the variance components  $\sigma^2$, and $\hat{g}_{k}$ is the estimated version (evaluated at the estimated values of the variance components $\hat{\sigma}^2$) of the gradient $g_{k}$ of $k'C_{11}k$, with respect to the variance components $\sigma^2$, i.e.
\begin{equation}\label{g1}
	g_{k} = \left(\begin{array}{c} 
	\frac{\partial \left(k'C_{11} k\right) }{\partial \sigma^2_1}\\
	\vdots\\
	\frac{\partial \left(k'C_{11} k\right) }{\partial \sigma^2_s}\\
	\frac{\partial \left(k'C_{11} k\right) }{\partial \sigma^2_{s+1}}\\
	\end{array} \right).
\end{equation}

As a generalization of the approach by Giesbrecht and Burns, it is natural to consider similar approximation for the distribution of the pivotal quantity (\ref{t3}), i.e.
\begin{equation}\label{distt3}
	t   = \frac{\hat{w}-w}{\sqrt{\lambda' \hat{C}\lambda}}  \sim t_{\hat{\nu}_{\lambda}},
\end{equation}
with
\begin{equation}\label{nu3}
\hat{\nu}_{\lambda}  = \frac{2 \left(\lambda'\hat{C} \lambda\right)^2}{\widehat{\var}(\lambda'\hat{C} \lambda)} 
\equiv  \frac{2 \left(\lambda'\hat{C} \lambda\right)^2}{\hat{g}_{\lambda}'\hat{\Sigma} \hat{g}_{\lambda}},
\end{equation}
where $\hat{g}_{\lambda}$ is the estimated version of the gradient $g_{\lambda}$ of $\mse\left(\tilde{w}\right) = \lambda' C\lambda $ with respect to the variance components $\sigma^2$, defined by
\begin{equation}\label{g3}
g_{\lambda} = \left(\begin{array}{c} 
\frac{\partial \left(\lambda'C\lambda \right) }{\partial \sigma^2_1}\\
\vdots\\
\frac{\partial \left(\lambda'C\lambda \right) }{\partial \sigma^2_s}\\
\frac{\partial \left(\lambda'C\lambda\right) }{\partial \sigma^2_{s+1}}\\
\end{array} \right).
\end{equation}
For more details on computing  gradients of the $\mse(\tilde{w})$ see Section~\ref{MSEderivatives}.

Provided that the estimated matrix $\hat{C}$ is available, e.g.~as an output of the algorithm for estimating the variance components, the estimators $\hat{g}_{k}$ and  $\hat{g}_{\lambda}$ of the gradients (\ref{g1}) and  (\ref{g3}) could be evaluated, by using the elements of the estimated matrix $\hat{C}$ (instead of $C$).

For that, let us define $\hat{\lambda} = \hat{C}\lambda$ and let $\hat{\lambda}$ be decomposed into its subvectors such that $\hat{\lambda} = (\hat{\lambda}_0', \hat{\lambda}_1' ,\dots,\hat{\lambda}_s')'$, where $\hat{\lambda}_0$ is $p$-dimensional subvector, and $\hat{\lambda}_i$, $i=1,\dots,s$, are $r_i$-dimensional subvectors of $\hat{\lambda}$.
Then, by using (\ref{d1MSE}) from Section~\ref{LMMM}, we get
\begin{equation}\label{hatg3}
	\hat{g}_{\lambda} = \left(
	\begin{array}{c} 
	\frac{1}{\left(\hat{\sigma}^2_{1}\right)^2} \hat{\lambda}_1'\hat{\lambda}_1\\
	\vdots\\
	\frac{1}{\left(\hat{\sigma}^2_{s}\right)^2} \hat{\lambda}_s'\hat{\lambda}_s  \\
	\frac{1}{\left(\hat{\sigma}^2_{s+1} \right)^2} \hat{\lambda}'H_0\hat{\lambda} \\
	\end{array} \right),
\end{equation}
where  $H_0  $ is given by  
\begin{equation}\label{H0}
H_0 = (X,Z)'(X,Z) = 
\left(\begin{array}{cc}
X'X & X'Z\cr
Z'X & Z'Z\end{array}\right).
\end{equation}

Consequently, as $k'b$ is a special case of $\lambda'\left(b',u'\right)' =k'b + l'u$ with $\lambda = \lambda_{(k)} = \left(k', 0_r'\right)'$, so we can use (\ref{hatg3}) also for evaluation of $\hat{g}_{k}$ by replacing $\hat{\lambda}$  with $\hat{\lambda}_{(k)} = \hat{C}\lambda_{(k)}$. 

\subsection{DDF estimated by the Fai-Cornelius method}\label{SectionFai}
Fai and Cornelius in \cite{Fai1996} proposed a generalization of the Satterthwaite's method for multivariate linear functions of the fixed and random effects  to approximate the (null) distribution of the statistic (\ref{F1}) by the Fisher-Snedecor $F$-distribution with $\nu_1 = q$ and $\nu_2 = \hat{\nu}$, i.e.~with the estimated denominator degrees of freedom (DDF).

As a straightforward generalization of the Fai-Cornelius approach,  it is natural to approximate the distribution of the $F$-statistic (\ref{F3}), based on the multivariate function $w= \Lambda'\left(b',u'\right)' = K'b + L'u$ and its empirical predictor $\hat{w} = K'\hat{b} + L'\hat{u}$, by the Fisher-Snedecor $F$-distribution with $\nu_1 = q$ and $\nu_2 = \hat{\nu}$ degrees of freedom, where
 where
\begin{equation}\label{hatnuFC}
	\hat{\nu}= \frac{2 \hat{E}}{\hat{E}-q},
\end{equation}
with
\begin{equation}\label{hatE}
	\hat{E} = \sum_{i=1}^q \frac{\hat{\nu}_i}{\hat{\nu}_i-2}       \mathop{\mathrm{1}}\nolimits_{\{\hat{\nu}_i > 2\}}   .
\end{equation}
Here, $\mathop{\mathrm{1}}\nolimits_{\{\cdot\}}$ denotes the indicator function and  $\hat{\nu}_i$, for $i = 1,\dots,q$, are the degrees of freedom, estimated by the Satterthwaite's method (\ref{nu3}), of the  $t$-statistics (\ref{t3}) for $\hat{w}_i = \hat{\lambda}_i' \left(\hat{b}',\hat{u}'\right)'$, where $\hat{\lambda}_i$, $i = 1,\dots,q$, are the columns of the matrix $\hat{\Lambda}_{FC}$ given by
\begin{equation}\label{HmatrixFC}
	\hat{\Lambda}_{FC} = \Lambda\hat{U},
\end{equation}
and $\hat{U}$ denotes the unitary matrix of a spectral decomposition of a matrix $\Lambda' \hat{C} \Lambda$, i.e.~such matrix that $\hat{U}' \Lambda' \hat{C} \Lambda \hat{U}=  \hat{S}$, where $\hat{S}$ is a diagonal matrix.
}

{
\section{Statistical inference on fixed and random effects based on adjusted estimator of the MSE matrix of the EBLUP}
As argued by Harville in \cite{Harville2008}, usage of the  MSE matrix of the BLUP $\tilde{w}$, say $M_{\tilde{w}}$, (or its estimated version, say $\widehat{M}_{\tilde{w}}$), instead of the correct MSE matrix of the EBLUP $\hat{w}$, say $M_{\hat{w}}$, (or its estimated version, say $\widehat{M}_{\hat{w}}$), is inadequate,  
as the estimator $\widehat{M}_{\tilde{w}} = \Lambda'\hat{C}\Lambda$ can severely underestimate the true MSE of the EBLUP $\hat{w}$. As will be explained bellow, there are two main sources of such bias.
For a comprehensive discussion on the problem and proposed solutions see also \cite{Kackar1981}, \cite{Kackar1984}, \cite{Harville1985}, \cite{Jeske1988}, \cite{Prasad1990}, \cite{Harville1992}, \cite{Jiang1999}, \cite{Schaalje2001}, \cite{Stulajter2002}, \cite{Stulajter2002a}, \cite{Das2004}, \cite{Kenward1997}, \cite{Kenward2009}, and \cite{Alnosaier2007}.

\subsection{Decomposition of the EBLUP prediction error and its MSE}
The first source of the bias can be observed if we decompose the prediction error of the EBLUP $\hat{w}$. In particular,
\begin{equation}\label{EBLUPerror}
	\left(\hat{w}-w \right) = \left(\tilde{w}-w \right) + \left(\hat{w}- \tilde{w} \right),
\end{equation}
and consequently, based on unbiasedness of EBLUP and its independence on BLUP, see \cite{Kackar1981}, \cite{Kackar1984}, \cite{Harville1985}, and \cite{Harville1992}, we get the MSE matrix of $\hat{w}$ in the  form
\begin{equation}\label{EBLUPmse}
	M_{\hat{w}} = M_{\tilde{w}} + M_{\delta\hat{w}},
\end{equation}
where $M_{\delta\hat{w}} = E\left(\left(\hat{w}- \tilde{w} \right) \left(\hat{w}- \tilde{w} \right)' \right) = \var\left(\hat{w}- \tilde{w} \right)$, and thus, $M_{\hat{w}} \geq M_{\tilde{w}}$.

The MSE of the first component of the prediction error, $M_{\tilde{w}}$, is given by (\ref{MSE}). The MSE of the second component of the prediction error, $M_{\delta\hat{w}}$, is not expressible in closed form, except for very simple special cases. Kackar and Harville in \cite{Kackar1984}, see also \cite{Kenward1997} and \cite{Kenward2009}, suggested approximation of $M_{\delta\hat{w}}$ based on first-order Taylor series approximation. 
In particular, a Taylor series expansion  for $\hat{w}-\tilde{w}$ in $\hat{\sigma}^2 = \left(\hat{\sigma}^2_1,\dots,\hat{\sigma}^2_s, \hat{\sigma}^2_{s+1}\right)'$, as e.g.~REML,
about $\sigma^2 = \left(\sigma^2_1,\dots,\sigma^2_s, \sigma^2_{s+1}\right)'$,  gives approximation
\begin{eqnarray}\label{Taylor1}
\left(\hat{w}- \tilde{w} \right) &\approx& \left(\tilde{w}- \tilde{w} \right) + \sum_{i=1}^{s+1}\frac{\partial \tilde{w}}{\partial \sigma^2_i} \left(\hat{\sigma}^2_i - \sigma^2_i\right) \cr
&& +\frac{1}{2} \sum_{i=1}^{s+1}\sum_{j=1}^{s+1} \frac{\partial^2 \tilde{w}}{\partial \sigma^2_i\sigma^2_j} \left(\hat{\sigma}^2_i - \sigma^2_i\right)\left(\hat{\sigma}^2_j-\sigma^2_j \right).
\end{eqnarray}
Then taking expectation of the square of the first-order term, and using the results in \cite{Kackar1984} and  \cite{Harville1992}, we get the first-order approximation $\dot{M}_{\delta\hat{w}} $ of $M_{\delta\hat{w}} $ as
\begin{eqnarray}\label{EBLUPmse2}
	\dot{M}_{\delta\hat{w}} &=& E\left( \frac{\partial \tilde{w}}{\partial \sigma^{2'}}\Sigma  \frac{\partial \tilde{w}'}{\partial \sigma^2}  \right)\cr 
	&=& \sum_{i=1}^{s+1}\sum_{j=1}^{s+1} \Sigma_{ij} E\left(\frac{\partial \tilde{w}}{\partial \sigma^2_i} \frac{\partial \tilde{w}'}{\partial \sigma^2_i}\right)\cr
	&=& \sum_{i=1}^{s+1}\sum_{j=1}^{s+1} \Sigma_{ij} \cov\left(\frac{\partial \left( \tilde{w}-w\right)}{\partial \sigma^2_i}, \frac{\partial \left(\tilde{w}-w\right)}{\partial \sigma^2_i}\right),
\end{eqnarray}
where $\Sigma_{ij}$ are elements of the variance-covariance matrix $\Sigma$ of the estimator  $\hat{\sigma}^2$. 

For derivation of the approximation of $\dot{M}_{\delta\hat{w}}$ see Section~\ref{LMMM2nd}.  
The second component of the EBLUP's MSE matrix $M_{\delta\hat{w}}$ in the simple LMM (\ref{model2}) can be approximated by
\begin{equation}\label{LMMM2main}
		\dot{M}_{\delta\hat{w}} = \sum_{i=1}^{s+1}\sum_{j=1}^{s+1} \Sigma_{ij} \mathbb{C}_{ij}.
\end{equation}
where $\mathbb{C}_{ij}$, $i,j = 1,\dots,s+1$, are given by (\ref{CCij}), or alternatively by
\begin{equation}\label{LMMM2mainb}
		\dot{M}_{\delta\hat{w}} = -\frac{1}{2}\sum_{i=1}^{s+1}\sum_{j=1}^{s+1} \Sigma_{ij} M_{\tilde{w}}^{(i,j)},
\end{equation}
where the matrices $M_{\tilde{w}}^{(i,j)}$ are given by (\ref{d2MSEii}), (\ref{d2MSEs1s1}), (\ref{d2MSEij}), and (\ref{d2MSEis1}). 

Consequently, we get the approximation $\dot{M}_{\hat{w}}$ of the EBLUP's MSE matrix $M_{\hat{w}}$ in the form
\begin{eqnarray}\label{LMMEBLUPMSE}
		\dot{M}_{\hat{w}} &=& M_{\tilde{w}}+\dot{M}_{\delta\hat{w}}\cr
&=& M_{\tilde{w}} + \sum_{i=1}^{s+1}\sum_{j=1}^{s+1} \Sigma_{ij} \mathbb{C}_{ij} \cr
&\equiv&	M_{\tilde{w}} -\frac{1}{2}\sum_{i=1}^{s+1}\sum_{j=1}^{s+1} \Sigma_{ij} M_{\tilde{w}}^{(i,j)},
\end{eqnarray}
where $\Sigma_{ij}$ are elements of the variance-covariance matrix of the REML estimator $\hat{\sigma}^2$, and $M_{\tilde{w}}^{(i,j)}$ represent the second partial derivatives of the BLUP's MSE matrix $M_{\tilde{w}}$ with respect to the variance components $\sigma^2_i$ and $\sigma^2_j$, $i,j=1,\dots,s+1$, in simple LMM (\ref{model2}).

\subsection{Bias-corrected estimator of the EBLUP's MSE matrix $M_{\hat{w}}$}
As the  EBLUP's MSE matrix $M_{\hat{w}}$, as well as its approximation $\dot{M}_{\hat{w}}$ (which is a function of $\Sigma$), depend on the unknown variance components $\sigma^2 = \left(\sigma^2_1,\dots,\sigma^2_{s+1}\right)'$, for further applications it is necessary to use its estimator, say $\widehat{\dot{M}}_{\hat{w}}$. A natural option for such estimator would be
\begin{equation}\label{estLMMEBLUPMSE}
	\widehat{\dot{M}}_{\hat{w}} =  \widehat{M}_{\tilde{w}}+\widehat{\dot{M}}_{\delta\hat{w}},
\end{equation}
i.e.~by using (\ref{LMMEBLUPMSE}), where the true (unknown) vector of variance components $\sigma^2$ is replaced by its estimator $\hat{\sigma}^2$. Notice that $\Sigma$, the true variance-covariance matrix of the REML estimator $\hat{\sigma}^2$ also depends on $\sigma^2$. So, the estimator (\ref{estLMMEBLUPMSE}) functionally depends on $\hat{\Sigma}_{ij}$, the elements of estimated variance-covariance matrix $\hat{\Sigma}$.

Based on similar arguments as given by Alnosaier in \cite{Alnosaier2007} for the special case of empirical BLUE of the  fixed effects, we can assume that $\widehat{\dot{M}}_{\delta\hat{w}}$ is approximately unbiased estimator of $M_{\delta\hat{w}}$,  for another formal justification see also \cite{Prasad1990} and \cite{Das2004}. 

However, as pointed out by Harville and Jeske in \cite{Harville1992}, Prasad and Rao in \cite{Prasad1990}, and in special case of fixed effects estimator by Kenward and Roger in \cite{Kenward1997} and \cite{Kenward2009}, additional bias will appear if the estimator $\widehat{M}_{\tilde{w}}$ is used as an estimators of the MSE matrix $M_{\tilde{w}}$ in (\ref{estLMMEBLUPMSE}). 
In order to show that,  let us expand $\widehat{M}_{\tilde{w}}$ in $\hat{\sigma}^2$ about $\sigma^2$, and then take expectation of this approximation, so
\begin{eqnarray}\label{Bias}
\lefteqn{\E\left(\widehat{M}_{\tilde{w}}\right) \approx M_{\tilde{w}} + \sum_{i=1}^{s+1} \E\left( \hat{\sigma}^2_i - \sigma^2_i\right) \frac{\partial M_{\tilde{w}}}{\partial \sigma^2_i}}\cr
&&\qquad + \frac{1}{2}\sum_{i=1}^{s+1}\sum_{j=1}^{s+1}\E\left(\left( \hat{\sigma}^2_i - \sigma^2_i\right)\left( \hat{\sigma}^2_j - \sigma^2_j\right)\right)
	 \frac{\partial^2 M_{\tilde{w}}}{\partial \sigma^2_i\partial \sigma^2_j}\cr
&&\ \quad \approx	 M_{\tilde{w}} + \frac{1}{2}\sum_{i=1}^{s+1}\sum_{j=1}^{s+1} \Sigma_{ij}M_{\tilde{w}}^{(i,j)}\cr
&&\ \quad =	 M_{\tilde{w}} - \dot{M}_{\delta\hat{w}},
\end{eqnarray}
where we have assumed that the first-order term could be ignored, and $\dot{M}_{\delta\hat{w}}$ is given by (\ref{LMMM2mainb}). This could be informally justified by the assumption that $\hat{\sigma}^2_i$ is approximately an unbiased estimator of $\sigma^2_i$, as was suggested in \cite{Kenward1997}. However,  formal justification was provided by Alnosaier in \cite{Alnosaier2007} and by Kenward and Roger in \cite{Kenward2009}. Kenward and Roger derived Taylor series approximation for the bias of REML estimator, i.e.~$\E\left( \hat{\sigma}^2_i - \sigma^2_i\right)$, and proved that in linear mixed models with linear parametrization of the variance-covariance matrix $V = Z'GZ+R$, like e.g.~in simple LMM (\ref{model2}), its first-order approximation is equal to zero.

Hence, by combining (\ref{estLMMEBLUPMSE}) and (\ref{Bias}), we get the adjusted, bias-corrected estimator of the EBLUP's MSE matrix $M_{\hat{w}}$,  given by
\begin{equation}\label{adjustedMSE}
\widehat{\dot{M}}_{\hat{w},A} =  \widehat{M}_{\tilde{w}}+2\widehat{\dot{M}}_{\delta\hat{w}}.
\end{equation}
The explicit form of the estimator (\ref{adjustedMSE}) in simple LMM (\ref{model2}) is given by (\ref{adjustedMSEform}) in Section~\ref{adjustedMSEsection}.

\subsection{Generalization of the  Kenward-Roger method for statistical inference on fixed and random effects based on adjusted estimator of the MSE matrix of the EBLUP}
For statistical inference about the vector of linear functions of fixed effects $K'b$ based on its empirical BLUE, Kenward and Roger suggested in \cite{Kenward1997} to use the Wald-type statistic as a pivot,  with adjusted covariance matrix of the empirical BLUE of the function $K'b$. 

Here we suggest to consider a generalization of the Kenward-Roger method for the inference about the vector of functions of fixed and random effects $w = \Lambda'(b', u')'$ (which is useful for testing hypotheses about the fixed effects and for constructing the prediction regions for functions of the fixed and the random effects simultaneously), based on its EBLUP and the adjusted MSE matrix. For that we shall consider the Wald-type pivot $F$-statistic
\begin{equation}\label{Fstat}
	F = \frac{1}{q}\left(\hat{w}-w \right)'\left( \widehat{\dot{M}}_{\hat{w},A}\right)^{-1}\left(\hat{w}-w \right),
\end{equation}
where $\widehat{\dot{M}}_{\hat{w},A}$ is given by (\ref{adjustedMSE}), or (in its explicit form) by (\ref{adjustedMSEform}) from Section~\ref{adjustedMSEsection}, respectively.

In accordance with \cite{Kenward1997} and \cite{Alnosaier2007}, we suggest to approximate the (null) distribution of the scaled Wald-type $F$-statistic (\ref{Fstat}) by the Fisher-Snedecor $F$-distribution with $q$ and $\nu$ degrees of freedom. In particular, 
\begin{equation}\label{Fstatdistribution}
\kappa	F \stackrel{\mathit{approx.}}{\sim} F_{q,\nu},
\end{equation}
where the unknown parameters $\kappa$ and $\nu$ should be estimated from the data.

In analogy with derivation of the estimators presented by Alnosaier in \cite{Alnosaier2007} for the fixed effects problem, here we suggest the following estimators of the scale $\kappa$ and the denominator degrees of freedom $\nu$:
\begin{eqnarray}\label{kappahat}
\hat{\kappa} &=& \frac{\hat{\nu}}{\hat{E}\left( \hat{\nu}-2\right)}, \cr
\hat{\nu} &=& 4 + \frac{2+q}{q\hat{\varrho} -1},
\end{eqnarray}
where
\begin{eqnarray}\label{details}
\hat{\varrho} &=& \frac{\hat{V}}{2\hat{E}^2},\cr
\hat{E} &=& 1 + \frac{\hat{A}_2}{q},\cr
\hat{V} &=& \frac{2}{q}\left(1+\hat{B} \right),\cr
\hat{B} &=& \frac{1}{2q}\left(\hat{A}_1 + 6\hat{A}_2 \right),
\end{eqnarray}
and
\begin{eqnarray}\label{details2}
\hat{A}_1 &=& \sum_{i=1}^{s+1} \sum_{j=1}^{s+1} \hat{\Sigma}_{ij}\tr\left(\widehat{M}_{\tilde{w}}^{-1} \widehat{M}_{\tilde{w}}^{(i)}\right)\tr\left(\widehat{M}_{\tilde{w}}^{-1} \widehat{M}_{\tilde{w}}^{(j)} \right),\cr
\hat{A}_2 &=& \sum_{i=1}^{s+1} \sum_{j=1}^{s+1} \hat{\Sigma}_{ij}\tr\left(\widehat{M}_{\tilde{w}}^{-1} \widehat{M}_{\tilde{w}}^{(i)}\widehat{M}_{\tilde{w}}^{-1} \widehat{M}_{\tilde{w}}^{(j)} \right).
\end{eqnarray}
By $\tr(A)$ we denote the trace of a matrix $A$, i.e. $\tr(A) =\sum_i\sum_j A_{ij}$,  $\widehat{M}_{\tilde{w}} = \Lambda'\hat{C}\Lambda$ denotes the estimated version of $M_{\tilde{w}}$, and 
$\widehat{M}_{\tilde{w}}^{(i)}$, $i=1,\dots,s+1$, denote the estimated versions of the first partial derivatives of $M_{\tilde{w}}$, defined by (\ref{d1MSE}). 
For more details and explicit forms of the estimators $\hat{A}_1$ and $\hat{A}_2$ see  Section~\ref{KRsection}, (\ref{A1}) and (\ref{A2}).

In order to match the exact values for the scale $\kappa$  and the denominator degrees of freedom $\nu$ for testing hypothesis on fixed effects in two special cases, in particular in the balanced one-way ANOVA and the Hotelling $T^2$ models, Kenward and Roger in \cite{Kenward1997} suggested the modified estimators $\hat{\kappa}^*$ and $\hat{\nu}^*$, which can be analogically generalized and  used to approximate the (null) distribution of the scaled Wald-type $F$-statistic (\ref{Fstat})
\begin{eqnarray}\label{kappahatmod}
\hat{\kappa}^* &=& \frac{\hat{\nu}^*}{\hat{E}^*\left( \hat{\nu}^*-2\right)}, \cr
\hat{\nu}^* &=& 4 + \frac{2+q}{q\hat{\varrho}^* -1},
\end{eqnarray}
where
\begin{eqnarray}\label{detailsmod}
\hat{\varrho}^* &=& \frac{\hat{V}^*}{2\hat{E}^{*2}},\cr
\hat{E}^* &=& \left(1 - \frac{\hat{A}_2}{q}\right)^{-1},\cr
\hat{V}^* &=& \frac{2}{q}\left(\frac{1+c_1 \hat{B}}{(1-c_2 \hat{B})^2(1-c_3 \hat{B})}\right),
\end{eqnarray}
and
\begin{eqnarray}\label{detailsmod2}
c_1 &=& \frac{g}{3q+2(1-g)},\cr
c_2 &=& \frac{q-g}{3q+2(1-g)},\cr
c_3 &=& \frac{q-g+2}{3q+2(1-g)},\cr
g &=& \frac{(q+1)\hat{A}_1-(q+4)\hat{A}_2}{(q+2)\hat{A}_2},
\end{eqnarray}
with $\hat{B}, \hat{A}_1, \hat{A}_2$ given by (\ref{details}) and (\ref{details2}). For more details see Section~4 in \cite{Alnosaier2007}.

\section{Conclusions}
Here we have presented a brief overview of the conventionally used methods for making statistical inference about linear functions of the fixed effects and/or about the fixed and random effects simultaneously, in conventional simple linear mixed model, by using the  elements of the solution of the Henderson's mixed model equations. 
Further, we have also presented some improvements, based on the adjusted MSE matrix of the EBLUP, as well as a generalization of the standard Kenward-Roger method (suggested for making statistical inference about the fixed effects) for derivation of the approximate distribution of the Wald-type pivot statistic, suggested for making statistical inference about the fixed and random effects simultaneously. 
Notice that this method for derivation of the approximate distribution of the Wald-type pivot statistic is not unique. As pointed out by Alnosaier in \cite{Alnosaier2007}, there are several other alternative solutions available, however, such modifications have not been considered here. 

The presented (explicit) expressions are valid in the simple LMM defined by (\ref{model2}). They are rather simple, and can be readily implemented in practically any (statistical) software environment. Based on the results presented in Section~\ref{MSEderivatives}, it is straightforward to get explicit expressions also for the more general LMM with linear parametrization of the variance-covariance matrices $G$ and $R$, provided that the REML of variance components and its estimated variance-covariance matrix is available. The situation with nonlinear parametrization of the matrices $G$ and $R$ requires more specific approach.

\section{Acknowledgements}
The work was supported by the Slovak Research and Development Agency, grant APVV-0096-10, and by the Scientific Grant Agency of the Ministry of Education of the Slovak Republic and the Slovak Academy of Sciences, grants VEGA 2/0038/12, 2/0019/10.

Valuable discussion and feedback from Barbora Arendack\'a, Francisco Carvalho, Augustyn Markiewicz, Jo{\~a}o T.~Mexia,  Roman Zmy\'{s}lony,  Tadeusz Cali\'{n}ski and Pawe{\l}  Krajewski, during the research group meeting on \emph{Sufficient and Optimal Statistical Procedures in Mixed Linear Model}, sponsored by the Stefan Banach International Mathematical Center, B\c{e}dlewo, Poland, November 11{--}17, 2012, is gratefully acknowledged, as well as  discussion on implementation of the Microsoft Excel version  of the algorithm which is under development by Mohammad Ovais of Xepa Soul Pattinson (Malaysia).
}

{
\appendix
\begin{center}
\textsc{Appendix}
\end{center}

\section{Derivatives of the MSE matrix with respect to the variance components}\label{MSEderivatives}
Here we shall assume that $G^{-1}$, the inverse of $G = \var(u)$, does exist, and thus we can use the MMEs as defined by (\ref{MME1}).
Although the subsequent derivation of the derivatives of the matrix $C$ is general, finally we shall consider only a special case, based on the covariance structure of the simple linear mixed model (\ref{model2}), with the variance-covariance matrices of the following form: $ G = \var(u) =\diag\{ \sigma^2_i I_{r_i}\}$, $i = 1,\dots,s$,  and $R = \var(e) = \sigma^2_{s+1} I_n$,  so $V = \var(y)= ZGZ'+ R = \sum_{i=1}^s\sigma^2_i Z_iZ_i'+\sigma^2_{s+1}I_n$.

Moreover, as we consider methods for statistical inference for estimable linear functions $w = \Lambda'(b', u')'= K'b + L'u$, i.e.~such that $K = X'A$ for some matrix $A$, further we shall assume, without loss of generality, that the inverse of the MME matrix $H$ (the matrix on the left-hand side of the equation (\ref{MME1})) does exist, in particular we shall assume that the inverse of $X'R^{-1}X$ does exist. Recall that 
\begin{equation}\label{2H}
H = (X,Z)'R^{-1}(X,Z) + (0,I_r)'G^{-1}(0,I_r),
\end{equation}
and so,
\begin{equation}\label{CH}
	C = H^{-1} \quad \mathrm{or} \quad  H = C^{-1},
\end{equation}
Further, we shall denote 
\begin{eqnarray}
\Delta_0  &=&(0,I_r)'(0,I_r),\label{D0}\\
\Delta_i &=&
\left(0,\left(0,\dots,I_{r_i},\dots,0\right)\right)'\left(0,\left(0,\dots,I_{r_i},\dots,0\right)\right)\cr
&=&
\left(\begin{array}{cc}
0 & 0\cr
0 & \diag_i\{I_{r_i}\} \end{array}\right),\label{Di}\\
\Delta_{s+1} &=& (X,Z)'(X,Z) = H_0,\label{Ds1}
\end{eqnarray}
for $i = 1,\dots,s$, where $\diag_i\{I_{r_i}\}$ is $(r\times r)$-matrix with its $i$-th diagonal block equal to $I_{r_i}$, otherwise with zero elements. 

Further, for arbitrary matrix $A$ we shall denote its partial derivatives with respect to the components of a vector parameter $\theta = \left(\theta_1,\dots,\theta_{s+1}\right)'$ as
\begin{equation}\label{d1H}
A^{(i)} = \frac{\partial A}{\partial \theta_i}, \ A^{(i,j)} = \frac{\partial^2 A}{\partial \theta_i \partial \theta_j}, \ 
A^{(i,j,k)} = \frac{\partial^3 A}{\partial \theta_i \partial \theta_j \partial \theta_k},
\end{equation}
for $i,j,k=1,\dots,s+1$.

Here we shall derive explicit expressions for derivatives of the matrix $C$, i.e.~$C^{(i)}$, $C^{(i,j)}$, and $C^{(i,j,k)}$, which depend on the derivatives of the matrices $G$ and $R$, i.e.~on $G^{(i)}$, $G^{(i,j)}$, $G^{(i,j,k)}$, and $R^{(i)}$, $R^{(i,j)}$, and $R^{(i,j,k)}$.

Recall that the derivative of $A^{-1}$, the inverse of a symmetric matrix $A$, with respect to some scalar parameter $\theta$, is given by
\begin{equation}\label{dAinv}
	\frac{\partial}{\partial \theta} A^{-1} = -A^{-1}\frac{\partial A}{\partial \theta} A^{-1},
\end{equation}
and the rule for computing the derivative of a symmetric matrix $ABA$ with respect to some parameter $\theta$ is
\begin{equation}\label{dABA}
	\frac{\partial}{\partial \theta} ABA 
	=  AB\frac{\partial A}{\partial \theta} + \frac{\partial A}{\partial \theta}BA +  A\frac{\partial B}{\partial \theta}A.
\end{equation}

Let $A$ be an inverse of a symmetric matrix $B$, i.e. $A = B^{-1}$. 
Then, based on (\ref{dAinv}) and (\ref{dABA}), we define the following matrix operators:
\begin{equation}\label{D1}
	{\cal D}^{(i)}\left(A,B\right) = - A B^{(i)} A,
\end{equation}
\begin{equation}\label{D2}
	{\cal D}^{(i,j)}\left(A,B\right) = A \left( B^{(i)}A B^{(j)} + B^{(j)}A B^{(i)} - B^{(i,j)}\right) A,	
\end{equation}
\begin{eqnarray}\label{D3}
\lefteqn{	{\cal D}^{(i,j,k)}\left(A,B\right) = -A \left( B^{(i)}A B^{(j)} + B^{(j)}A B^{(i)} - B^{(i,j)}\right) A B^{(k)}A}\cr
&& -A B^{(k)}A \left( B^{(i)}A B^{(j)} + B^{(j)}A B^{(i)} - B^{(i,j)}\right) A\cr
&& + A \left( B^{(i)}A B^{(j,k)} + B^{(i,k)}A B^{(j)} + B^{(j)}A B^{(i,k)} + B^{(j,k)}A B^{(i)}-  \right. \cr
&& \left.  - B^{(i)}A B^{(k)}A B^{(j)} - B^{(j)}A B^{(k)}A B^{(i)}A + B^{(i,j,k)}\right) A.
\end{eqnarray}

From that we directly get
\begin{eqnarray}
C^{(i)} &=& {\cal D}^{(i)}\left(C,H\right) \cr
&=& - C H^{(i)} C,\label{d1C2}\\
	C^{(i,j)} &=& {\cal D}^{(i,j)}\left(C,H\right)\cr
&=& C \left( H^{(i)}C H^{(j)} +  H^{(j)}C H^{(i)} - H^{(i,j)}\right) C,	\label{d2C2}\\
C^{(i,j,k)} &=& {\cal D}^{(i,j,k)}\left(C,H\right),\label{d3C2}
\end{eqnarray}
for $i,j,k = 1,\dots,s+1$, 

Further, based on (\ref{2H}), we directly get the derivatives of the matrix $H$. 
For $i,j,k = 1,\dots,s$ 
\begin{eqnarray}
H^{(i)} &=& (0,I_r)' G^{-1 (i)}(0,I_r),\label{d1H2i}\\
H^{(s+1)} &=& (X, Z)'R^{-1 (s+1)} (X,Z),\label{d1H2s1}\\
H^{(i,j)} &=& (0,I_r)' G^{-1 (i,j)}(0,I_r),\label{d2H2i}\\
H^{(s+1,s+1)} &=& (X, Z)'R^{-1 (s+1,s+1)} (X,Z),\label{d2H2s1}\\
H^{(i,j,k)} &=& (0,I_r)' G^{-1 (i,j,k)}(0,I_r),\label{d3H2i}\\
H^{(s+1,s+1,s+1)} &=& (X, Z)'R^{-1 (s+1,s+1,s+1)} (X,Z),\label{d3H2s1}
\end{eqnarray}
where 
\begin{eqnarray}
G^{-1 (i)} &=& {\cal D}^{(i)}\left( G^{-1},G \right)\cr
G^{-1 (i,j)} &=& {\cal D}^{(i,j)}\left( G^{-1},G \right)\cr
G^{-1 (i,j,k)} &=& {\cal D}^{(i,j,k)}\left( G^{-1},G \right)\cr
R^{-1 (s+1)} &=& {\cal D}^{(s+1)}\left( R^{-1},R \right)\cr
R^{-1 (s+1,s+1)} &=&  {\cal D}^{(s+1,s+1)}\left( R^{-1},R \right)\cr
R^{-1 (s+1,s+1,s+1)} &=&  {\cal D}^{(s+1,s+1,s+1)}\left( R^{-1},R \right).
\end{eqnarray}
Notice that 
\begin{eqnarray}\label{d2H2is1}
	H^{(i,s+1)} &=& H^{(s+1,j)} = 0,\quad i,j\neq s+1 \\
	H^{(i,j,k)} &=& 0,\label{d3H2is1}
\end{eqnarray}
whenever one index is equal to $s+1$ and some of the other indices is different from for $s+1$, for $i,j,k = 1,\dots,s+1$.

\subsection{Derivatives of the MME matrix $H$ in simple LMM}\label{LMMH}
In the simple LMM (\ref{model2}), we get
\begin{eqnarray}
H^{(i)} &=& -\frac{1}{\left(\sigma^2_i \right)^2} \Delta_i,\label{d1H2iG}\\
H^{(i,i)} &=& \frac{2}{\left(\sigma^2_i \right)^3} \Delta_i,\label{d2H2iiG}\\
H^{(i,i,i)} &=& -\frac{6}{\left(\sigma^2_i \right)^4} \Delta_i,\label{d3H2iiiG}
\end{eqnarray}
for $i=1,\dots,s+1$. Notice that 

\begin{equation}\label{d2H2iG}
	H^{(i,j)} = 0, \quad \mathrm{and} \quad H^{(i,j,k)} = 0,
\end{equation}
for any combination of unequal indices $i,j,k = 1,\dots,s+1$. 

\subsection{Derivatives of the MME matrix $C$ in simple LMM}\label{LMMC}
By combining (\ref{d1C2}), (\ref{d2C2}), (\ref{d1H2iG}), (\ref{d2H2iiG}), and (\ref{d2H2iG}), in simple LMM (\ref{model2}), we directly get
\begin{eqnarray}
C^{(i)} &=&  \frac{1}{\left(\sigma^2_i \right)^2} C\Delta_i C, \label{d1C3}\\
C^{(i,i)} &=& \frac{2}{\left(\sigma^2_{i} \right)^4} C\left( \Delta_iC \Delta_i  - \sigma^2_{i}\Delta_i\right)C, \label{d2C4}\\
C^{(i,j)} &=& \frac{1}{\left(\sigma^2_{i}\sigma^2_{j}  \right)^2} C\left( \Delta_iC \Delta_j + \Delta_jC \Delta_i\right)C, \  i\neq j, \label{d2C3}
\end{eqnarray}
for $i,j = 1,\dots,s+1$.

The explicit expression for  $C^{(i,j,k)}$, i.e.~the third partial derivative of $C$ for $i,j,k = 1,\dots,s+1$, is not presented here, however, it can be similarly evaluated based on (\ref{d3C2}),
(\ref{d1H2iG}), (\ref{d2H2iiG}), (\ref{d3H2iiiG}), and (\ref{d2H2iG}).

\subsection{Derivatives of the MSE matrix $M_{\tilde{w}}$ in simple LMM}\label{LMMM}
Recall that $M_{\tilde{w}}$, the MSE matrix of the best linear unbiased predictor of $w$, is given by $M_{\tilde{w}}= \Lambda'C\Lambda$, where $\Lambda$ is $((p+r)\times q)$-matrix of given coefficients. 

Let $\tilde{\Lambda}$ be a solution of a system of linear equations  $H \tilde{\Lambda}= \Lambda$, i.e.~$\tilde{\Lambda} = C\Lambda$, and let $\tilde{\Lambda}$ be decomposed into block-matrices such that $\tilde{\Lambda} = (\tilde{\Lambda}_0', \tilde{\Lambda}_1' ,\dots,\tilde{\Lambda}_s')'$, where $\tilde{\Lambda}_0$ is $(p\times q)$-dimensional block-matrix, and $\tilde{\Lambda}_i$, $i=1,\dots,s$, are $(r_i\times q)$-dimensional block-matrices of $\tilde{\Lambda}$. 
Similarly, let $\{ C\}_{ij}$ denote the $(i,j)$-th block\footnote{Notice that for $i,j=1,\dots,s$ the block $\{ C\}_{ij} = \{ C_{22}\}_{ij}$, i.e.~it is the $(i,j)$-th block of the matrix $C_{22}$, which can be, based on (\ref{C}), efficiently computed as $C_{22} = \sigma^2_{s+1}G\left(\sigma^2_{s+1}I_r + MG \right)^{-1}$, where $M = Z'Z-Z'X(X'X)^{-}X'Z$.}  
of the matrix $C$, and let $\{ C\}_{i\cdot}$ denote the $i$-th row-block and $\{ C\}_{\cdot i}$ the $i$-th column-block of the matrix $C$.
 
Then,  based on the derivatives of the matrix $C$, we directly get the first partial derivatives of the MSE matrix $M_{\tilde{w}}$ with respect to the variance components
$\sigma^2_1,\dots,\sigma^2_s,\sigma^2_{s+1}$ as
\begin{eqnarray}\label{d1MSE}
M_{\tilde{w}}^{(i)} &=&  \frac{1}{\left(\sigma^2_{i} \right)^2} \tilde{\Lambda}'\Delta_i\tilde{\Lambda} =  \frac{1}{\left(\sigma^2_{i} \right)^2}\tilde{\Lambda}_i'\tilde{\Lambda}_i,\ i = 1,\dots,s,\cr
M_{\tilde{w}}^{(s+1)} &=&  \frac{1}{\left(\sigma^2_{s+1} \right)^2} \tilde{\Lambda}'\Delta_{s+1}\tilde{\Lambda},
\end{eqnarray}
where the matrices $\Delta_i$ are defined by (\ref{Di}) and (\ref{Ds1}). 
The second  partial derivatives of $M_{\tilde{w}}$ are given by:
\begin{eqnarray}\label{d2MSEii}
M_{\tilde{w}}^{(i,i)} &=& \frac{2}{\left(\sigma^2_{i} \right)^4} \tilde{\Lambda}'\left( \Delta_iC\Delta_i   - \sigma^2_{i} \Delta_i\right)\tilde{\Lambda} \cr
&=& \frac{2}{\left(\sigma^2_{i} \right)^4} \left(\tilde{\Lambda}_i'\{C\}_{ii}\tilde{\Lambda}_i   - \sigma^2_{i} \tilde{\Lambda}_i'\tilde{\Lambda}_i\right),
\end{eqnarray}
for $i = 1,\dots,s$, and in for $i = s+1$ we get
\begin{equation}\label{d2MSEs1s1}
M_{\tilde{w}}^{(s+1,s+1)}
= \frac{2}{\left(\sigma^2_{s+1} \right)^4} \tilde{\Lambda}'\left(\Delta_{s+1}C\Delta_{s+1}   - \sigma^2_{s+1}\Delta_{s+1}\right)\tilde{\Lambda},
\end{equation}
Further,
\begin{eqnarray}\label{d2MSEij}
M_{\tilde{w}}^{(i,j)} &=&  M_{\tilde{w}}^{(j,i)}\cr
&=& \frac{1}{\left(\sigma^2_{i}\sigma^2_{j}  \right)^2} \tilde{\Lambda}'\left(  \Delta_iC\Delta_j + \Delta_jC\Delta_i \right)\tilde{\Lambda} \cr
 &=& \frac{1}{\left(\sigma^2_{i}\sigma^2_{j}  \right)^2}  \left( \tilde{\Lambda}_i'\{C\}_{ij}\tilde{\Lambda}_j + \tilde{\Lambda}_j'\{C\}_{ji}\tilde{\Lambda}_i\right),
\end{eqnarray}
for $i\neq j$, $i,j = 1,\dots,s$, and
\begin{eqnarray}\label{d2MSEis1}
M_{\tilde{w}}^{(i,s+1)}&=&  M_{\tilde{w}}^{(s+1,i)} \cr
&=& \frac{1}{\left(\sigma^2_{i}\sigma^2_{s+1}  \right)^2} 
\tilde{\Lambda}'\left( \Delta_iC \Delta_{s+1}  + \Delta_{s+1}C\Delta_i\right)\tilde{\Lambda}, \cr
&=& \frac{1}{\left(\sigma^2_{i}\sigma^2_{s+1} \right)^2}  
\left( \tilde{\Lambda}_i'\{C\}_{i\cdot} \Delta_{s+1}\tilde{\Lambda} \right. \cr
&& \qquad \qquad  \qquad  +   \left. \tilde{\Lambda}' \Delta_{s+1} \{C\}_{\cdot i} \tilde{\Lambda}_i\right),
\end{eqnarray}
for $i = 1,\dots,s$.
}

\subsection{Approximation of the second component of the EBLUP's MSE matrix in simple LMM}\label{LMMM2nd}
According to (\ref{EBLUPmse2}), let us define $\dot{M}_{\delta\hat{w}}$ by
\begin{eqnarray}\label{defEBLUPmse2}
	\dot{M}_{\delta\hat{w}} 
	&=& \sum_{i=1}^{s+1}\sum_{j=1}^{s+1} \Sigma_{ij} \cov\left(\frac{\partial \left( \tilde{w}-w\right)}{\partial \sigma^2_i}, \frac{\partial \left(\tilde{w}-w\right)}{\partial \sigma^2_i}\right),\cr
	&=& \sum_{i=1}^{s+1}\sum_{j=1}^{s+1} \Sigma_{ij} \mathbb{C}_{ij}
\end{eqnarray}
where $\Sigma_{ij}$ denote the elements of the variance-covariance matrix $\Sigma$ of $\hat{\sigma}^2$. Then, by using 
\begin{equation}
	\tilde{w}-w = \Lambda' C (X,Z)'R^{-1}\left(y - Xb \right) - \Lambda'(0,I_r)'u,
\end{equation}
we get
\begin{eqnarray}
\frac{\partial \left(\tilde{w}-w\right)}{\partial \sigma^2_i} &=& -\Lambda'CH^{(i)}C(X,Z)'R^{-1}(y-Xb)\cr
&& - \Lambda'C (X,Z)'R^{-1}R^{(i)}R^{-1}(y-Xb).
\end{eqnarray}
and then, by taking the covariances of the vectors with $i,j = 1,\dots,s+1$, we get, 
\begin{eqnarray}\label{bbC}
\lefteqn{\mathbb{C}_{ij} = \Lambda' C \Bigg( H^{(i)}C(X,Z)' R^{-1} V R^{-1}(X,Z)C H^{(j)}} \cr
&& + \  H^{(i)}C(X,Z)' R^{-1} V R^{-1}R^{(j)} R^{-1}(X,Z) \cr
&& + \ (X,Z)' R^{-1} R^{(i)} R^{-1}V R^{-1}(X,Z)C H^{(j)}\cr
&& + \  C(X,Z)' R^{-1} R^{(i)} R^{-1} V R^{-1}R^{(j)} R^{-1}(X,Z)\Bigg)C\Lambda,\cr
&& \ \hfill
\end{eqnarray}
where $V = ZGZ' +R$. 

Notice that in the simple LMM (\ref{model2}) we have $R^{(i)} = R^{(j)} = 0$, for $i,j=1,\dots,s$, and $R^{(s+1)} = I_n$. From that we get $R^{-1} R^{(s+1)} = R^{(s+1)} R^{-1} = R^{-1} = \frac{1}{\sigma^2_{s+1}} I_n$, and 
\begin{eqnarray}\label{CCij}
\lefteqn{\mathbb{C}_{i,j} = \frac{1}{\left(\sigma^2_{i}\sigma^2_{j}\right)^2} \Lambda'C\Big(\Delta_{i}CH_VC\Delta_{j}} \cr
&&  -1_{\{i=s+1\}} \sigma^2_{s+1}H_VC\Delta_{j} -1_{\{j=s+1\}}\sigma^2_{s+1}\Delta_{i}CH_V  \cr
&&+ 1_{\{i=j=s+1\}}\left(\sigma^2_{s+1}\right)^2 H_V \Big)C\Lambda,
\end{eqnarray}
for $i,j=1,\dots,s+1$, where $1_{\{i=s+1\}}$, $1_{\{j=s+1\}}$, $1_{\{i=j=s+1\}}$ are the indicator functions, and $H_V = (X,Z)' R^{-1} V R^{-1}(X,Z)$  fulfills the  property  
\begin{equation}\label{CHVC}
	CH_VC = \left(\begin{array}{cc} C_{11} & 0 \cr 0 & G-C_{22}\end{array}\right). 
\end{equation}
Hence, the approximation of the second component of the EBLUP's MSE matrix, i.e.~$\dot{M}_{\delta\hat{w}}$, in simple LMM is 
\begin{equation}\label{LMMM2}
\dot{M}_{\delta\hat{w}} = \sum_{i=1}^{s+1}\sum_{j=1}^{s+1} \Sigma_{ij} \mathbb{C}_{ij}.
\end{equation}
with $\mathbb{C}_{ij}$, $i,j = 1,\dots,s+1$, given by (\ref{CCij}). 

By recognizing that in simple LMM (\ref{model2}) we have $M_{\tilde{w}}^{(i,i)} = -2\mathbb{C}_{i,i}$ and $M_{\tilde{w}}^{(i,j)} = -\left( \mathbb{C}_{i,j}+\mathbb{C}_{j,i}\right) $, $i,j=1,\dots,s+1$, see also \cite{Harville2008} eq.~(4.6), we get the alternative expression for the
approximation of the second component of the EBLUP's MSE matrix in simple LMM, given by
\begin{equation}\label{LMMM2b}
		\dot{M}_{\delta\hat{w}} = -\frac{1}{2}\sum_{i=1}^{s+1}\sum_{j=1}^{s+1} \Sigma_{ij} M_{\tilde{w}}^{(i,j)},
\end{equation}
where the matrices $M_{\tilde{w}}^{(i,j)}$ are given by (\ref{d2MSEii}), (\ref{d2MSEs1s1}), (\ref{d2MSEij}), and (\ref{d2MSEis1}).

{
\subsection{Bias-corrected estimator of the MSE matrix of EBLUP in simple LMM}\label{adjustedMSEsection}
In simple LMM (\ref{model2}), the bias-corrected estimator of the MSE matrix of the empirical BLUP of $w = \Lambda'(b',u')'$, i.e.~$M_{\hat{w}}$, is given (based on (\ref{adjustedMSE}) and (\ref{LMMM2b})), as 
\begin{eqnarray}\label{Mhatw}
\widehat{\dot{M}}_{\hat{w},A}  &=& \widehat{M}_{\tilde{w}} + 2\widehat{\dot{M}}_{\delta\hat{w}}\cr
&=& \widehat{M}_{\tilde{w}} - \left(\sum_{i=1}^{s+1} \sum_{j=1}^{s+1} \hat{\Sigma}_{ij} \widehat{M}_{\tilde{w}}^{(i,j)}\right),
\end{eqnarray}
and in particular, by using $\widehat{M}_{\tilde{w}} = \Lambda'\hat{C}\Lambda$ and (\ref{d2MSEii}), (\ref{d2MSEs1s1}), (\ref{d2MSEij}), and (\ref{d2MSEis1}), we get
\begin{eqnarray}\label{adjustedMSEform}
\lefteqn{\widehat{\dot{M}}_{\hat{w},A} = \Lambda'\hat{\Lambda} + \frac{4\hat{\Sigma}_{s+1,s+1}}{\left(\hat{\sigma}^2_{s+1} \right)^4} \hat{\Lambda}'\left(\hat{\sigma}^2_{s+1}H_0 - H_0 \hat{C}H_0 \right)\hat{\Lambda}}\cr
&+& \sum_{i=1}^{s}  \frac{4\hat{\Sigma}_{ii}}{\left(\hat{\sigma}^2_{i}  \right)^4} \left(\hat{\sigma}^2_{i}\hat{\Lambda}_i'\hat{\Lambda}_i- \hat{\Lambda}_i'\{\hat{C}\}_{ii}\hat{\Lambda}_i\right)\cr
&-& \sum_{i=1}^{s} \frac{4\hat{\Sigma}_{i,s+1}}{\left(\hat{\sigma}^2_{i}\hat{\sigma}^2_{s+1}  \right)^2}  \left( \hat{\Lambda}_i'\{\hat{C}\}_{i\cdot} H_0 \hat{\Lambda} + \hat{\Lambda}'H_0  \{\hat{C}\}_{\cdot i}\hat{\Lambda}_i\right) \cr
&-& \mathop{\sum\sum}_{i<j}^{s} \frac{4\hat{\Sigma}_{ij}}{\left(\hat{\sigma}^2_{i}\hat{\sigma}^2_{j}  \right)^2}  \left( \hat{\Lambda}_i'\{\hat{C}\}_{ij}\hat{\Lambda}_j+ \hat{\Lambda}_j'\{\hat{C}\}_{ji}\hat{\Lambda}_i\right) ,
\end{eqnarray}
where $\hat{\Lambda} = \hat{C}\Lambda$,  $H_0 = \Delta_{s+1} = (X,Z)'(X,Z)$, and  $\hat{\Sigma}$, (with elements $\hat{\Sigma}_{ij}$, $i,j=1,\dots,s+1$), is the estimated
variance-covariance matrix of the REML estimator $\hat{\sigma}^2 = \left(\hat{\sigma}^2_1,\dots, \hat{\sigma}^2_{s+1}\right)'$.
Here, $\hat{\Lambda} = \left(\hat{\Lambda}_0', \hat{\Lambda}_1' ,\dots,\hat{\Lambda}_s'\right)'$ is decomposed into block-matrices such that $\hat{\Lambda}_0$ is $(p\times q)$-dimensional block-matrix, and $\hat{\Lambda}_i$, $i=1,\dots,s$, are $(r_i\times q)$-dimensional block-matrices of $\hat{\Lambda}$. 
Similarly, $\{ \hat{C}\}_{ij}$ denote the $(i,j)$-th  $(r_i\times r_j)$-dimensional block of the matrix $\hat{C}$, and  $\{ \hat{C}\}_{i\cdot}$ denote the $i$-th $(r_i\times(p+r))$-dimesional row-block and $\{ \hat{C}\}_{\cdot i}$ the $i$-th $((p+r)\times r_i)$-dimesional column-block of the matrix $\hat{C}$.

\subsection{Generalized Kenward-Roger method for statistical inference on fixed and random effects based on adjusted estimator of the MSE matrix of the EBLUP in simple LMM}\label{KRsection}
Here we shall consider the scaled Wald-type  $F$-statistic defined by (\ref{Fstat}), in particular
\begin{equation}\label{Fstat2}
	\kappa F_* = \frac{\kappa}{q}\left(\hat{w}-w \right)'\left( \widehat{\dot{M}}_{\hat{w},A}\right)^{-1}\left(\hat{w}-w \right) \stackrel{\mathit{approx.}}{\sim} F_{q,\nu},
\end{equation}
where $\widehat{\dot{M}}_{\hat{w},A}$ is given by (\ref{adjustedMSEform}).

The moment based estimators of the parameters $\kappa$ and $\nu$ are based on comparing the first and the second moments of the scaled $F$-statistic (\ref{Fstat2}) with the moments of the  $F$-distribution  with $q$ and $\nu$ degrees of freedom, i.e.~by solving the system of equations 
\begin{eqnarray}
{\E}(\kappa F_*) = {\kappa}{E_*}  &\cong&  E = \E(F_{q,{\nu}}),\cr
{\var}(\kappa F_*) = {\kappa}^2{V_*} &\cong& V = \var(F_{q,{\nu}}),
\end{eqnarray}
where ${E_*} = \E(F_*)$ and ${V_*} = \var(F_*)$. 
Based on the properties of the $F$-distribution we get
\begin{eqnarray}
E &=& \frac{\nu}{\nu-2},\cr
V &=& \frac{2\nu^2(\nu+q-2)}{q(\nu-2)^2(\nu-4)} \cr
&=& \frac{2E^2}{q} \frac{\nu+q-2}{\nu-4},
\end{eqnarray}
provided that $\nu>4$. By denoting 
\begin{equation}\label{varrho}
	\varrho = \frac{V}{2E^2} 
\end{equation}
we get
\begin{equation}
	\nu = 4+\frac{q+2}{q\varrho-1},
\end{equation}
and consequently, the moment estimators of $\kappa$ and $\nu$ are given as
\begin{eqnarray}\label{nutilde}
\tilde{\kappa} &=& \frac{\tilde{\nu}}{{E_*}(\tilde{\nu}-2)}\cr
\tilde{\nu} &=& 4+\frac{q+2}{q\tilde{\varrho}-1},
\end{eqnarray}
where
\begin{equation}
\tilde{\varrho} = \frac{{V_*}}{2{E_*}^2}.
\end{equation}
The expectation and the variance of the statistic $F_*$ defined by (\ref{Fstat2}) can be estimated by using
\begin{eqnarray}
E_* &=& \E\left(F_*\right) = \E_{\hat{\sigma}^2}\left(\E_{\hat{w}}\left(F_* \,|\, \hat{\sigma}^2\right) \right) \cr
V_* &=& \var\left(F_*\right) = \E_{\hat{\sigma}^2}\left(\var_{\hat{w}}\left(F_* \,|\, \hat{\sigma}^2\right) \right) \cr
&& +\var_{\hat{\sigma}^2}\left( \E_{\hat{w}}\left(F_*\,|\, \hat{\sigma}^2\right)\right).
\end{eqnarray}
Alnosaier in \cite{Alnosaier2007} derived approximations for $E_*$ and $V_*$ in the special case, when the $F$-statistic (\ref{Fstat2}) is restricted on fixed effects only. 
The derivation of the approximations $E_*$ and $V_*$ in the general case, (i.e.~for the $F$-statistic defined by (\ref{Fstat2})), is not presented here. 
However, in analogy with the derivation of the approximations presented in \cite{Alnosaier2007}, we suggest $\dot{E}_*$ and $\dot{V}_*$, as the  approximations of $E_*$ and $V_*$, in the following form
\begin{eqnarray}\label{Adetails}
\dot{E}_* &=& 1 + \frac{A_2}{q},\cr
\dot{V}_* &=& \frac{2}{q}\left(1+B \right),
\end{eqnarray}
where
\begin{eqnarray}\label{Adetails2}
B &=& \frac{1}{2q}\left(A_1 + 6A_2 \right),\cr
A_1 &=& \sum_{i=1}^{s+1} \sum_{j=1}^{s+1} \Sigma_{ij}\tr\left(M_{\tilde{w}}^{-1} M_{\tilde{w}}^{(i)}\right)
\tr\left(M_{\tilde{w}}^{-1} M_{\tilde{w}}^{(j)} \right),\cr
A_2 &=& \sum_{i=1}^{s+1} \sum_{j=1}^{s+1} \Sigma_{ij}
\tr\left(M_{\tilde{w}}^{-1} M_{\tilde{w}}^{(i)}M_{\tilde{w}}^{-1} M_{\tilde{w}}^{(j)} \right).
\end{eqnarray}

The suggested approximations depend on the unknown variance components $\sigma^2 = \left(\sigma^2_1,\dots,\sigma^2_{s+1}\right)'$. 
Consequently, the suggested estimators of the parameters $\kappa$ and $\nu$,  based on the estimated versions of (\ref{nutilde}), are
\begin{eqnarray}\label{nuhat}
\hat{\kappa} &=& \frac{\hat{\nu}}{\widehat{\dot{E}}_*(\hat{\nu}-2)}\cr
\hat{\nu} &=& 4+\frac{q+2}{q\hat{\varrho}-1},
\end{eqnarray}
where
\begin{equation}
\hat{\varrho} = \frac{{\widehat{\dot{V}}_*}}{2{\widehat{\dot{E}}_*}^2},
\end{equation}
and
\begin{eqnarray}\label{hatdetails}
\widehat{\dot{E}}_* &=& 1 + \frac{\hat{A}_2}{q},\cr
\widehat{\dot{V}}_* &=& \frac{2}{q}\left(1+\hat{B} \right),
\end{eqnarray}
with
\begin{eqnarray}\label{hatdetails2}
\hat{B} &=& \frac{1}{2q}\left(\hat{A}_1 + 6\hat{A}_2 \right),\cr
\hat{A}_1 &=& \sum_{i=1}^{s+1} \sum_{j=1}^{s+1} \hat{\Sigma}_{ij}\tr\left(\widehat{M}_{\tilde{w}}^{-1} \widehat{M}_{\tilde{w}}^{(i)}\right)
\tr\left(\widehat{M}_{\tilde{w}}^{-1} \widehat{M}_{\tilde{w}}^{(j)} \right),\cr
\hat{A}_2 &=& \sum_{i=1}^{s+1} \sum_{j=1}^{s+1} \hat{\Sigma}_{ij}
\tr\left(\widehat{M}_{\tilde{w}}^{-1} \widehat{M}_{\tilde{w}}^{(i)}\widehat{M}_{\tilde{w}}^{-1} \widehat{M}_{\tilde{w}}^{(j)} \right).
\end{eqnarray}
In particular, by using $\widehat{M}_{\tilde{w}} = \Lambda'\hat{C}\Lambda = \Lambda'\hat{\Lambda}$ and (\ref{d1MSE}), we finally get
\begin{eqnarray}\label{A1}
\hat{A}_1 &=& 
\sum_{i=1}^{s}  \frac{\hat{\Sigma}_{ii}}{\left(\hat{\sigma}^2_i \right)^4} \tr\left(\left(\Lambda'\hat{\Lambda} \right)^{-1} \hat{\Lambda}_i'\hat{\Lambda}_i \right)^2\cr
&&+\mathop{\sum\sum}_{i<j}^{s} \frac{2\hat{\Sigma}_{ij}}{\left(\hat{\sigma}^2_i \hat{\sigma}^2_j\right)^2} \cr
&&\quad \times
\tr\left(\left(\Lambda'\hat{\Lambda} \right)^{-1} \hat{\Lambda}_i'\hat{\Lambda}_i \right)\tr\left(\left(\Lambda'\hat{\Lambda} \right)^{-1} \hat{\Lambda}_j'\hat{\Lambda}_j \right)\cr
&& + \sum_{i=1}^{s} \frac{\hat{2\Sigma}_{i,s+1}}{\left(\hat{\sigma}^2_i\hat{\sigma}^2_{s+1}\right)^2} \cr
&&\quad \times
\tr\left(\left(\Lambda'\hat{\Lambda} \right)^{-1} \hat{\Lambda}_i'\hat{\Lambda}_i \right)\tr\left(\left(\Lambda'\hat{\Lambda} \right)^{-1} \hat{\Lambda}'H_0\hat{\Lambda} \right)\cr
&&+\frac{\hat{\Sigma}_{s+1,s+1}}{\left(\hat{\sigma}^2_{s+1}\right)^4} 
\tr\left(\left(\Lambda'\hat{\Lambda} \right)^{-1} \hat{\Lambda}'H_0\hat{\Lambda} \right)^2,
\end{eqnarray}
\begin{eqnarray}\label{A2}
\hat{A}_2 &=& 
\sum_{i=1}^{s}  \frac{\hat{\Sigma}_{ii}}{\left(\hat{\sigma}^2_i \right)^4} \tr\left(\left(\left(\Lambda'\hat{\Lambda} \right)^{-1} \hat{\Lambda}_i'\hat{\Lambda}_i \right)^2\right)\cr
&&+\mathop{\sum\sum}_{i<j}^{s} \frac{2\hat{\Sigma}_{ij}}{\left(\hat{\sigma}^2_i \hat{\sigma}^2_j\right)^2} \cr
&&\quad \times
\tr\left(\left(\Lambda'\hat{\Lambda} \right)^{-1} \hat{\Lambda}_i'\hat{\Lambda}_i\left(\Lambda'\hat{\Lambda} \right)^{-1} \hat{\Lambda}_j'\hat{\Lambda}_j \right)\cr
&& + \sum_{i=1}^{s} \frac{\hat{2\Sigma}_{i,s+1}}{\left(\hat{\sigma}^2_i\hat{\sigma}^2_{s+1}\right)^2} \cr
&&\quad \times
\tr\left(\left(\Lambda'\hat{\Lambda} \right)^{-1} \hat{\Lambda}_i'\hat{\Lambda}_i\left(\Lambda'\hat{\Lambda} \right)^{-1} \hat{\Lambda}'H_0\hat{\Lambda} \right)\cr
&&+\frac{\hat{\Sigma}_{s+1,s+1}}{\left(\hat{\sigma}^2_{s+1}\right)^4} 
\tr\left(\left(\left(\Lambda'\hat{\Lambda} \right)^{-1} \hat{\Lambda}'H_0\hat{\Lambda} \right)^2\right),
\end{eqnarray}
as before, $\hat{\Lambda} = \hat{C}\Lambda$,  $H_0 = \Delta_{s+1} = (X,Z)'(X,Z)$, and  $\hat{\Sigma}$, (with elements $\hat{\Sigma}_{ij}$, $i,j=1,\dots,s+1$), is the estimated
variance-covariance matrix of the REML estimator $\hat{\sigma}^2 = \left(\hat{\sigma}^2_1,\dots, \hat{\sigma}^2_{s+1}\right)'$. 
$\hat{\Lambda} = \left(\hat{\Lambda}_0', \hat{\Lambda}_1' ,\dots,\hat{\Lambda}_s'\right)'$ is decomposed into block-matrices such that $\hat{\Lambda}_0$ is $(p\times q)$-dimensional block-matrix, and $\hat{\Lambda}_i$, $i=1,\dots,s$, are $(r_i\times q)$-dimensional block-matrices of $\hat{\Lambda}$. 
Similarly, $\{ \hat{C}\}_{ij}$ denote the $(i,j)$-th  $(r_i\times r_j)$-dimensional block of the matrix $\hat{C}$, and  $\{ \hat{C}\}_{i\cdot}$ denote the $i$-th $(r_i\times(p+r))$-dimensional row-block and $\{ \hat{C}\}_{\cdot i}$ the $i$-th $((p+r)\times r_i)$-dimensional column-block of the matrix $\hat{C}$.
}

{
\section{Estimation of the variance components by solving the MMEs}\label{Estimation}
The presented iterative procedure for estimation of the variance components by solving the Henderson's mixed model equations has been suggested by Searle, Casella and McCulloch in \cite{Searle1992},  see pp.~275--286. The MATLAB version of the algorithm has been implemented by Witkovsk\'y in \cite{Witkovsky2001}.

Here we use the same notation as in \cite{Searle1992}.
In each step of the suggested iterative procedure, we shall denote $V^{(t)}=\sigma^{2(t)}_{s+1}I_r+Z'ZG^{(t)}$,
$G^{(t)}=\diag\left(\sigma^{2(t)}_iI_{r_i}\right)$. The algorithm starts with the choice of the starting values for variance components $\sigma^{2(0)}=\left(\sigma^{2(0)}_1, \dots, \sigma^{2(0)}_{s+1}\right)'$
and setting $t=0$.
In the $t$-th step of the procedure the algorithm solves the system of mixed model equations:
\begin{equation}
\left(\begin{array}{cc}
X'X & X'ZG^{(t)}\cr
Z'X & V^{(t)}\end{array}\right)
\left(\begin{array}{c}\tilde{b}^{(t)} \cr
\tilde{v}^{(t)} \end{array} \right)=
\left(\begin{array}{c}X'y\cr Z'y \end{array} \right),
\end{equation}
and $\tilde{u}^{(t)}=G^{(t)}\tilde{v}^{(t)}$.

\subsection{ML estimates of the variance components}
The ML estimates of the variance components are calculated iteratively as 
\begin{eqnarray}
{\sigma^2_i}^{(t+1)} &=&
\frac{ \tilde{u}_i^{(t)'} \tilde{u}_i^{(t)}}
{{r}_i-\tr\left(W_{ii}^{(t)}\right)}, \quad i=1,\dots,s,\cr
{\sigma^2_{s+1}}^{(t+1)} &=&
\frac{y'\left(y-X\tilde{b}^{(t)}-Z\tilde{u}^{(t)}\right)}{n},
\end{eqnarray}
where $\tilde{u}_i^{(t)}$ is the $i$-th $r_i$-dimensional subvector of $\tilde{u}^{(t)}$
and $W_{ii}^{(t)}$ is the $i$-th diagonal block of the matrix
$W^{(t)}$, where
\begin{equation}
W^{(t)}=\sigma^{2(t)}_{s+1} {V^{(t)}}^{-1}
=\sigma^{2(t)}_{s+1}
\left(\sigma^{2(t)}_{s+1}I_{r}+Z'ZG^{(t)}\right)^{-1}.
\end{equation}

The iterative procedure should be stopped after the $t$-th step if $\left\|{\sigma^2}^{(t)}- {\sigma^2}^{(t-1)}\right\|<\varepsilon$,
for the chosen precision limit $\varepsilon$, and where
${\sigma^2}^{(t)}= \left({\sigma^2_{1}}^{(t)},\dots,{\sigma^2_{r+1}}^{(t)}\right)'$.

The final solutions of the iterative procedure are denoted by
$\hat{b}$, $\hat{u}=\left(\hat{u}_1',\dots,\hat{u}_{s}'\right)'$,
and $\hat{\sigma}^2= \left(\hat{\sigma}^2_1,\dots,\hat{\sigma}^2_{s+1}\right)'$.
Similarly, we denote $\hat{W}$ and use the adequate  notation $\hat{G}$, $\hat{R}$,  and $\hat{C}$ for the estimated versions of matrices $G$, $R$,  and $C$.

The log-likelihood function for ML estimation evaluated at the ML estimates $\hat{b}$ and $\hat{\sigma}^2$, say $\loglik_{ML}$, is
\begin{eqnarray}
\loglik_{ML} &=& -\frac{1}{2}n\log(2\pi)
-\frac{1}{2}\log\left(|\hat{V}|\right)\cr
&&-\frac{1}{2}\left(y-X\hat{b}\right)'{\hat{V}}^{-1}\left(y-X\hat{b}\right),\cr
&=& -\frac{1}{2}\left(
n\log\left(2\pi\hat{\sigma}^2_{s+1}\right) -\log\left(|\hat{W}|\right)+n \right),
\end{eqnarray}
where $\hat{V}=Z\hat{G}Z'+\hat{\sigma}^2_{s+1} I_{n}$
and $\hat{W} = \left(I_{r}+Z'Z\hat{G}/ \hat{\sigma}^2_{s+1} \right)^{-1}$.

The Fisher information matrix (which is in fact the inverse of the asymptotic variance-covariance matrix) of the ML estimators of the variance
components, say ${I}_{ML}({\sigma}^2)$,
can be evaluated at the ML estimates
$\hat{\sigma}^2$ as
{\small
\begin{eqnarray}\label{IML}
\lefteqn{{I}_{ML}\left(\hat{\sigma}^2\right)=\frac{1}{2}\times}\cr
&& 
\hspace{-23pt}\left(\begin{array}{cc}
\Bigl\{_{mat}\!\!\!\!\!\!\!\!\!\!
\frac{\delta_{ij}[{r}_i-2\tr(\hat{W}_{ii})]
+\tr(\hat{W}_{ij}\hat{W}_{ji})}
{\hat{\sigma}^2_i\hat{\sigma}^2_j}\Bigr\}_{i,j=1}^{s} &
\Bigl\{_{col}\!\!\!\!\!\!
\frac{\tr(\hat{W}_{ii})-
\sum_{j}^s\tr(\hat{W}_{ij}\hat{W}_{ji})}
{\hat{\sigma}^2_i\hat{\sigma}^2_{s+1}}\Bigr\}_{i=1}^{s}\cr
\Bigl\{_{row}\!\!\!\!\!\!\!\!
\frac{\tr(\hat{W}_{ii})
-\sum_{j}^s\tr(\hat{W}_{ij}\hat{W}_{ji})}
{\hat{\sigma}^2_i\hat{\sigma}^2_{s+1}}\Bigr\}_{i=1}^{s}&
\frac{n-{m}+\tr(\hat{W}^2)}{\hat{\sigma}^4_{s+1}}
\end{array} \right),\, \quad\mbox{ }
\end{eqnarray}}
where $\delta_{ij}=1$ if $i=j$, otherwise  $\delta_{ij}=0$, and $\hat{W}_{ij}$ is the $(r_i\times r_j)$ block of the matrix $\hat{W}$.

\subsection{REML estimates of the variance components}
Similarly, the REML estimates of the variance components are calculated iteratively as
\begin{eqnarray}
{\sigma^2_i}^{(t+1)} &=&
\frac{ \tilde{u}_i^{(t)'} \tilde{u}_i^{(t)}}
{{r}_i-\tr\left(T_{ii}^{(t)}\right)},  \quad i=1,\dots,s,\cr
{\sigma^2_{s+1}}^{(t+1)} &=&
\frac{y'\left(y-X\tilde{b}^{(t)}-Z\tilde{u}^{(t)}\right)}{n-r_X},
\end{eqnarray}
where by $r_X$ we denote the rank of the matrix $X$,
$\tilde{u}_i^{(t)}$ is the $i$-th $r_i$-dimensional
subvector of $\tilde{u}^{(t)}$ and
$T_{ii}^{(t)}$ is the $i$-th diagonal block of the matrix
$T^{(t)}$, where
\begin{equation}
T^{(t)}=
{\sigma^2_{s+1}}^{(t)}
\left( {\sigma^2_{s+1}}^{(t)} I_{r}+ MG^{(t)}\right)^{-1},
\end{equation}
where $M=Z' Z -Z'X(X'X)^{-}X'Z$.

The log-likelihood function for REML estimation evaluated at the REML estimates $\hat{\sigma}^2$, say $\loglik_{REML}$, is
\begin{eqnarray}
\loglik_{REML} &=& -\frac{1}{2}\left(n-r_X\right)\log(2\pi)
-\frac{1}{2}\log\left(|B'\hat{V}B|\right)\cr
&&-\frac{1}{2}y'B(B'\hat{V}B)^{-1}B'y,\cr
&=& -\frac{1}{2} (n-r_X)\log\left(2\pi\hat{\sigma}^2_{s+1}\right)\cr
&& -\frac{1}{2}\left(
-\log\left(|\hat{T}|\right)+(n-r_X) \right),
\end{eqnarray}
where $B$ is an $n\times(n-r_X)$ matrix, such that $BB'=I_{n}-X(X'X)^{-}X'$ and $B'B=I_{n-r_X}$. Further, $\hat{T}=\left(I_{r}+M\hat{G}/ \hat{\sigma}^2_{s+1} \right)^{-1}$.

The Fisher information matrix of the REML estimators of the variance components, ${I}_{REML}({\sigma}^2)$,
can be evaluated at the REML estimates
$\hat{\sigma}^2$ as
{\small \begin{eqnarray}\label{IREML}
\lefteqn{{I}_{REML}\left(\hat{\sigma}^2\right)=\frac{1}{2}\times}\cr
&&\hspace{-23pt}\left(\begin{array}{cc}
\Bigl\{_{mat}\!\!\!\!\!\!\!\!
\frac{\delta_{ij}[{r}_i-2\tr(\hat{T}_{ii})]
+\tr(\hat{T}_{ij}\hat{T}_{ji})}
{\hat{\sigma}^2_i\hat{\sigma}^2_j}\Bigr\}_{i,j=1}^{s} &
\Bigl\{_{col}\!\!\!\!\!\!
\frac{\tr(\hat{T}_{ii})
-\sum_{j}^s\tr(\hat{T}_{ij}\hat{T}_{ji})}
{\hat{\sigma}^2_i\hat{\sigma}^2_{s+1}}\Bigr\}_{i=1}^{s}\cr
\Bigl\{_{row}\!\!\!\!\!\!\!\!
\frac{\tr(\hat{T}_{ii})
-\sum_{j}^s\tr(\hat{T}_{ij}\hat{T}_{ji})}
{\hat{\sigma}^2_i\hat{\sigma}^2_{s+1}}\Bigr\}_{i=1}^{s}&
\frac{n-r_X-r+\tr(\hat{T}^2)}{\hat{\sigma}^4_{s+1}}
\end{array} \right),\ \quad\mbox{ }
\end{eqnarray}}
where $\delta_{ij}=1$ if $i=j$, otherwise  $\delta_{ij}=0$, and
$\hat{T}_{ij}$ is the $(r_i\times r_j)$ block of the matrix $\hat{T}$. 

Similarly, the final solutions of the procedure are denoted by
$\hat{b}$, $\hat{u}=\left(\hat{u}_1',\dots,\hat{u}_{s}'\right)'$, and $\hat{\sigma}^2= \left(\hat{\sigma}^2_1,\dots,\hat{\sigma}^2_{s+1}\right)'$. Further,
we denote $\hat{T}$, and use the adequate  notation $\hat{G}$, $\hat{R}$,  and $\hat{C}$ for the estimated versions of matrices $G$, $R$,  and $C$.

For more details on ML and REML estimators see the Chapter~6 in Searle {et al.} (1992).

\subsection{MINQE's of the variance components}
For completeness, here we present procedures to calculate the MINQE(I) and the
MINQE(U,I) estimators of the variance components at given (prior)
values of the variance components
$\sigma^{2(0)}=\left(\sigma^{2(0)}_1, \dots, \sigma^{2(0)}_{s+1}\right)'$.
Here we assume that $\sigma^{2(0)}_i>0$ for all $i=1,\dots,s+1$.
For more details on minimum norm quadratic estimation of the
variance components see e.g.~\cite{LaMotte1973}, \cite{Rao1972}, and \cite{Kleffe1988}.

The MINQE(I) of $\sigma^2$, say $\hat{\sigma}^2$,
at the prior value $\sigma^{2(0)}$ is defined as
the solution of the following system of equations
\begin{equation}\label{minqei}
H_{(I)}\hat{\sigma}^2=q,
\end{equation}
where by $H_{(I)}$ we denote the $(s+1 \times s+1)$-dimensional MINQE(I)-matrix and
$q=\left(q_1,\dots,q_{s+1}\right)'$ denotes the vector of MINQE quadratic forms.
The matrix $H_{(I)}$ is defined by its elements as
\begin{equation}
\left\{H_{(I)} \right\}_{ij}
=\tr\left(V^{(0)^{-1}}V_iV^{(0)^{-1}}V_j\right),
\end{equation}
$i,j=1,\dots,s+1$, where $V_i = Z_iZ_i'$, for $i=1,\dots,s$,
$V_{s+1}=I_{n}$,  and
$V^{(0)}=ZG^{(0)}Z'+\sigma^{2(0)}_{s+1}I_{n}=
\sum_{i=1}^{s+1}\sigma^{2(0)}_iV_i$.
The matrix $H_{(I)}$ can be easily evaluated by using (\ref{IML}), namely
\begin{equation}
H_{(I)} = 2 {I}_{ML}\left({\sigma}^{2(0)}\right).
\end{equation}

Further, the vector $q$ of MINQE
quadratic forms, defined by its elements as
\begin{equation}
q_i= y'\left(M_XV^{(0)}M_X\right)^{+}V_i\left(M_XV^{(0)}M_X\right)^{+}y,
\end{equation}
$i=1,\dots,s+1$, with $M_X=I_{n}-X(X'X)^{-}X$, could be easily evaluated by using
\begin{eqnarray}
q_i &=&
\frac{ \tilde{u}_i^{(0)'} \tilde{u}_i^{(0)}}
{\left(\sigma^{2(0)}_i\right)^2}, \quad i=1,\dots,s,\cr
q_{s+1} &=&
\frac{\left(y-X\tilde{b}^{(0)}-Z\tilde{u}^{(0)}\right)'
\left(y-X\tilde{b}^{(0)}-Z\tilde{u}^{(0)}\right)}
{\left(\sigma^{2(0)}_{s+1}\right)^2},
\end{eqnarray}
where $\tilde{u}_i^{(0)}$ is the $i$-th $r_i$-dimensional
subvector of $\tilde{u}^{(0)}$.

Similarly, the MINQE(U,I) of $\sigma^2$, say $\hat{\sigma}^2$,
at the prior value $\sigma^{2(0)}$
is defined as the solution of the following system of equations
\begin{equation}\label{minqeui}
H_{(UI)}\hat{\sigma}^2=q,
\end{equation}
where $H_{(UI)}$ denotes the $(s+1 \times s+1)$-dimensional MINQE(U,I) matrix, defined by its elements
\begin{equation}
\left\{H_{(UI)} \right\}_{ij}
=\tr\left(\left(M_XV^{(0)}M_X\right)^{+}V_i\left(M_XV^{(0)}M_X\right)^{+}V_j\right),
\end{equation}
$i,j=1,\dots,s+1$, and by using (\ref{IREML}) we get
\begin{equation}
H_{(UI)} = 2 {I}_{REML}\left({\sigma}^{2(0)}\right).
\end{equation}

Note that the MINQE $\hat{\sigma}^2$, defined by (\ref{minqei}) or by (\ref{minqeui}), is not given uniquely unless the MINQE matrix is of full rank.
In fact, one version of the solution to the MINQE equations is $\hat{\sigma}^2=H^+q$, where $H^+$ denote the Moore-Penrose $g$-inverse of the appropriate MINQE matrix.

The MINQE of unbiasedly estimable vector $F{\sigma}^2$, where $F$ is such matrix that $F'=HA$ for some matrix $A$, is $F\hat{\sigma}^2$, and is unique.

In particular, under given assumptions, the MINQE(U,I) $F\hat{\sigma}^2$, with $F$ such that $F'=H_{(UI)}A$ for some matrix $A$,
is  the
$\sigma^{2(0)}$-locally minimum variance unbiased invariant
estimator of $F{\sigma}^2$ with
\begin{eqnarray}
\E\left(F\hat{\sigma}^2\right) &=& F{\sigma}^2, \cr
\var\left(F\hat{\sigma}^2 \,|\, \sigma^{2(0)}\right) &=& 2 FH^-_{(UI)}F'\cr
&=&2A'H_{(UI)}A.
\end{eqnarray}

On the other hand, the MINQE(I) $F\hat{\sigma}^2$ 
is a biased estimator of $F{\sigma}^2$ with
\begin{eqnarray}
\E(F\hat{\sigma}^2) &=& FH_{(I)}^-H_{(UI)}{\sigma}^2, \cr
\var\left(F\tilde{\sigma}^2 \,|\, \sigma^{2(0)}\right) &=& 2
FH_{(I)}^-H_{(UI)}H_{(I)}^-F'.
\end{eqnarray}
}

\end{document}